\documentclass[bibyear]{aa}  

\usepackage{graphicx}
\usepackage{txfonts}
\usepackage{natbib}
\bibpunct{(}{)}{;}{a}{}{,} 
\usepackage{booktabs} 
\newcommand{\ra}[1]{\renewcommand{\arraystretch}{#1}} 
\usepackage{natbib,twoopt}
\usepackage[breaklinks=true]{hyperref} 
\bibpunct{(}{)}{;}{a}{}{,}             
\makeatletter
  \newcommandtwoopt{\citeads}[3][][]{\href{http://adsabs.harvard.edu/abs/#3}%
    {\def\hyper@linkstart##1##2{}%
     \let\hyper@linkend\@empty\citealp[#1][#2]{#3}}}
  \newcommandtwoopt{\citepads}[3][][]{\href{http://adsabs.harvard.edu/abs/#3}%
    {\def\hyper@linkstart##1##2{}%
     \let\hyper@linkend\@empty\citep[#1][#2]{#3}}}
  \newcommandtwoopt{\citetads}[3][][]{\href{http://adsabs.harvard.edu/abs/#3}%
    {\def\hyper@linkstart##1##2{}%
     \let\hyper@linkend\@empty\citet[#1][#2]{#3}}}
  \newcommandtwoopt{\citeyearads}[3][][]%
    {\href{http://adsabs.harvard.edu/abs/#3}
    {\def\hyper@linkstart##1##2{}%
     \let\hyper@linkend\@empty\citeyear[#1][#2]{#3}}}
\makeatother
\usepackage{orcidlink}

\begin{document}

   \title{Radial Evolution of ICME-Associated Particle Acceleration Observed by Solar Orbiter and ACE}


   \author{Malik H. Walker \orcidlink{0000-0003-4380-3519}
          \inst{1}
          \and
          Robert C. Allen \orcidlink{0000-0003-2079-5683} \inst{2}
          \and
          Gang Li \orcidlink{0000-0003-4695-8866} \inst{3}
          \and
          George C. Ho \orcidlink{0000-0003-1093-2066}\inst{2}
          \and
          Glenn M. Mason \orcidlink{0000-0003-2169-9618} \inst{4}
          \and
          Javier Rodriguez-Pacheco \orcidlink{0000-0002-4240-1115} \inst{5}
          \and
          Robert F. Wimmer-Schweingruber \orcidlink{0000-0002-7388-173X} \inst{6}
          \and
          Athanasios Kouloumvakos \orcidlink{0000-0001-6589-4509} \inst{4}
          }

   \institute{Johns Hopkins University,
             Baltimore, MD, USA 21218\\
              \email{mwalke85@jhu.edu}
        \and
            Southwest Research Institute, San Antonio, TX, USA 78238
        \and 
            General Linear Space Plasma Lab LLC, Hunstville, AL, USA 35899
         \and
             Johns Hopkins Applied Physics Laboratory, Laurel, MD, USA 20723-6099
        \and 
            University of Alcalá, Alcalá de Henares, Madrid, Spain
        \and
            University of Kiel, Kiel, Germany}

   \date{Received XXXX; accepted XXXX}

 
  \abstract
  {On 2022 March 10, a coronal mass ejection (CME) erupted from the Sun, resulting in Solar Orbiter observations at 0.45 au of both dispersive solar energetic particles arriving prior to the interplanetary CME (ICME) and locally accelerated particles near the ICME-associated shock structure as it passed the spacecraft on 2022 March 11. This shock was later detected on 2022 March 14 by the Advanced Composition Explorer (ACE), which was radially aligned with Solar Orbiter, at 1 au. Ion composition data from both spacecraft— via the Solar Orbiter Energetic Particle Detector/ Suprathermal Ion Spectrograph (EPD/SIS) and the Ultra Low Energy Isotope Spectrometer (ULEIS) on ACE— allows for in-depth analysis of the radial evolution of species-dependent ICME shock-associated acceleration processes for this event. We present a study of the ion spectra observed at 0.45 and 1 au during both the gradual solar energetic particle (SEP) and energetic storm particle (ESP) phases of the event. We find that the shapes of the spectra seen at each spacecraft have significant differences that were likely caused by varying shock geometry: Solar Orbiter spectra tend to lack spectral breaks, and the higher energy portions of the ACE spectra have comparable average flux to the Solar Orbiter spectra. Through an analysis of rigidity effects on the spectral breaks observed by ACE, we conclude that the 1 au observations were largely influenced by a suprathermal pool of $\mathrm{He}^{+}$ ions that were enhanced due to propagation along a stream interaction region (SIR) that was interacting with the ICME at times of observation.}

   \keywords{Sun: coronal mass ejections (CMEs) --
                shock waves --
                solar wind
               }

   \maketitle
%

\section{Introduction}

Solar energetic particles (SEPs), high energy particles accelerated by the Sun in eruptive events, are an important part of the heliospheric environment. SEP events can vary significantly in both composition and size, and are generally separated into two categories: impulsive and gradual. While impulsive SEP events originating from solar flares occur more frequently, gradual SEP events tend to have greater intensity and longer duration. Gradual SEPs originate from shocks driven by coronal mass ejections (CMEs) expanding into interplanetary space, and are accelerated near the corona during the onset of the shock formation. Often lasting for days, gradual SEPs contain particles with energies ranging from $\sim$ 30 keV to a few GeV, with the highest energy particles leading to ground level enhancements (GLEs) on rare occasions \citep{mew2001,chen2022}. For fast CMEs with well developed shocks directed towards an observing spacecraft, these initial SEPs are typically followed by an energetic storm particle (ESP) event: a large increase in low energy particles accelerated locally at the interplanetary CME (ICME) - driven shock, and associated with its crossing \citep{giac2020}. Since particles from both gradual SEP and ESP events can pose a serious threat to microelectronics on satellites, as well as humans both on high-latitude aircraft flights and in space, they are of particular importance to the field of space weather (\citet{tsu2003,garcia2016}, for a review on SEPs, see \citet{cohen2021}). 

th gradual SEPs and ESPs is widely thought to be diffusive shock acceleration (DSA, for a review see \citet{dru1983}). Within this process, particles gain energy by interacting with irregularities in the magnetic field present at the shock. Diffusive effects allow particles to scatter between the converging upstream and downstream regions of the plasma that bound the shock front, accelerating them to higher energies. The DSA mechanism states that the particles accelerated within the shock should have a power-law momentum distribution, $f \propto p^{-\eta}$, where $\eta$ is solely dependent on the ratio of the downstream to upstream plasma density, or shock compression ratio, and the slope of the power law is determined by the shock kinematics \citep{ty2000}. However, in practice the particle spectra observed for large SEP events are often better fit by either the Ellison-Ramaty (ER) function \citep{er1985} or a variation of the double power law \citep{band1993}. These profiles contain spectral breaks that are directly proportional to the charge-to-mass ratio of the ion and vary based on species if the diffusion coefficient at the shock has the form $\kappa \sim \beta \mathrm{R}^{\alpha}$ (where $\beta$ is the particle speed, $\mathrm{R}$ is the magnetic rigidity) and $\alpha = 1$. The ER function also has an exponential rollover at high energies due to finite shock-lifetime and/or finite shock-size effects \citep{ty2000,chen2022}. Although a number of theories have been proposed to explain the applicability of the ER function and double power law in fitting spectra from large SEP and ESP events, there is still a significant amount uncertainty regarding the effects of shock evolution on the shape of these spectra.

ring their transit, ICMEs can also interact with other large scale structures in the solar wind, a process that can alter the geometry and trajectories of the coronal ejecta (e.g., \citet{luhm2020,shen2017,shen2018,giac2021}), ultimately affecting the efficiency of particle acceleration and transport at the associated shock \citep{tsu2003,chen2022,wij2021}. One such structure is a stream interaction region (SIR), an area of compressed plasma formed when fast solar wind streams from coronal holes expand into the preceding slow solar wind. These events can accelerate particles at forward and reverse shock pairs that develop along their leading edges, typically forming at heliospheric distances beyond 1 au \citep{rich2004}. However, once accelerated, some of these particles will propagate back into the inner heliosphere. This can be seen in the consistent observations of energetic particle intensity peaks increasing as a function of radial distance across ion species within 1 au (e.g., \citet{allen2021, vanh1978}). The accelerated particles from SIRs can then contribute to the suprathermal seed population \citep{wij2023} which can be further accelerated at ICME-associated shocks. As a result, ICME/SIR interactions can vary the elemental compositions of energetic particles through changes in the suprathermal pool as well as affect the efficiency of shock associated particle acceleration through changes to the shock geometry. An improved understanding of these effects is necessary to develop more robust SEP prediction models for future events.

bservations are one of the key methods of analyzing SEP events, especially with regard to their lonitudinal spread. A number of such studies have been done, such as \citet{cohen2017}, which used the STEREO spacecraft situated at 1 au together with ACE to analyze the variations in particle composition and spectra as a function of longitudinal separation without the influence of radial dependency. While these studies are useful in providing constraints for SEP acceleration and transport models between the Sun and 1 au, in situ measurements near the Sun, where a majority of the particle acceleration is occurring, are required to fully deduce the actual conditions within the region. With the current wealth of satellites distributed within the inner and outer heliosphere, in situ observations of ICMEs at different stages of their evolution are more accessible than ever (e.g., for enhanced longitudinal studies: \citet{koll2021,mason2021}, for studies focused on radial variations: \citet{palm2022, palm2024, rod2023}). Furthermore, the analysis of measurements taken for specific ion species is also a notable tool for probing transport and acceleration processes, due to the particle velocity and the mass-per-charge ratio dependence of particle transport equations \citep{lee2000}.

On 2022 March 10, an ICME was detected by Solar Orbiter at 0.45 au, with both dispersive SEP onset signatures and locally accelerated particles near the ICME-associated shock structure. This shock was later detected on 2022 March 14 by ACE at 1 au, when ACE was nearly radially aligned with Solar Orbiter (within $7.2^{\circ}$ longitude and $2.9^{\circ}$ latitude). Given this specific configuration, this event provides a prime opportunity to better understand the radial dependency of ICME shock-associated particle acceleration. Additionally, simulations show that a SIR is expected to be continuously interacting with the ICME during its propagation to 1 au. We present analysis of the radial evolution of ICME-associated species-dependent acceleration processes from 0.45 to 1 au for the 2022 March 10 ICME. Section \ref{Obs} provides an overview of the event, Section \ref{Results} presents the SEP and ESP spectra. A discussion of the results is given in Section \ref{Disc}, and the conclusions are given in Section \ref{Conc}.

\section{Observations}\label{Obs}
\subsection{Instrumentation}\label{inst}
 \begin{figure*}[h!]
   \centering
            {\includegraphics[width = \hsize]{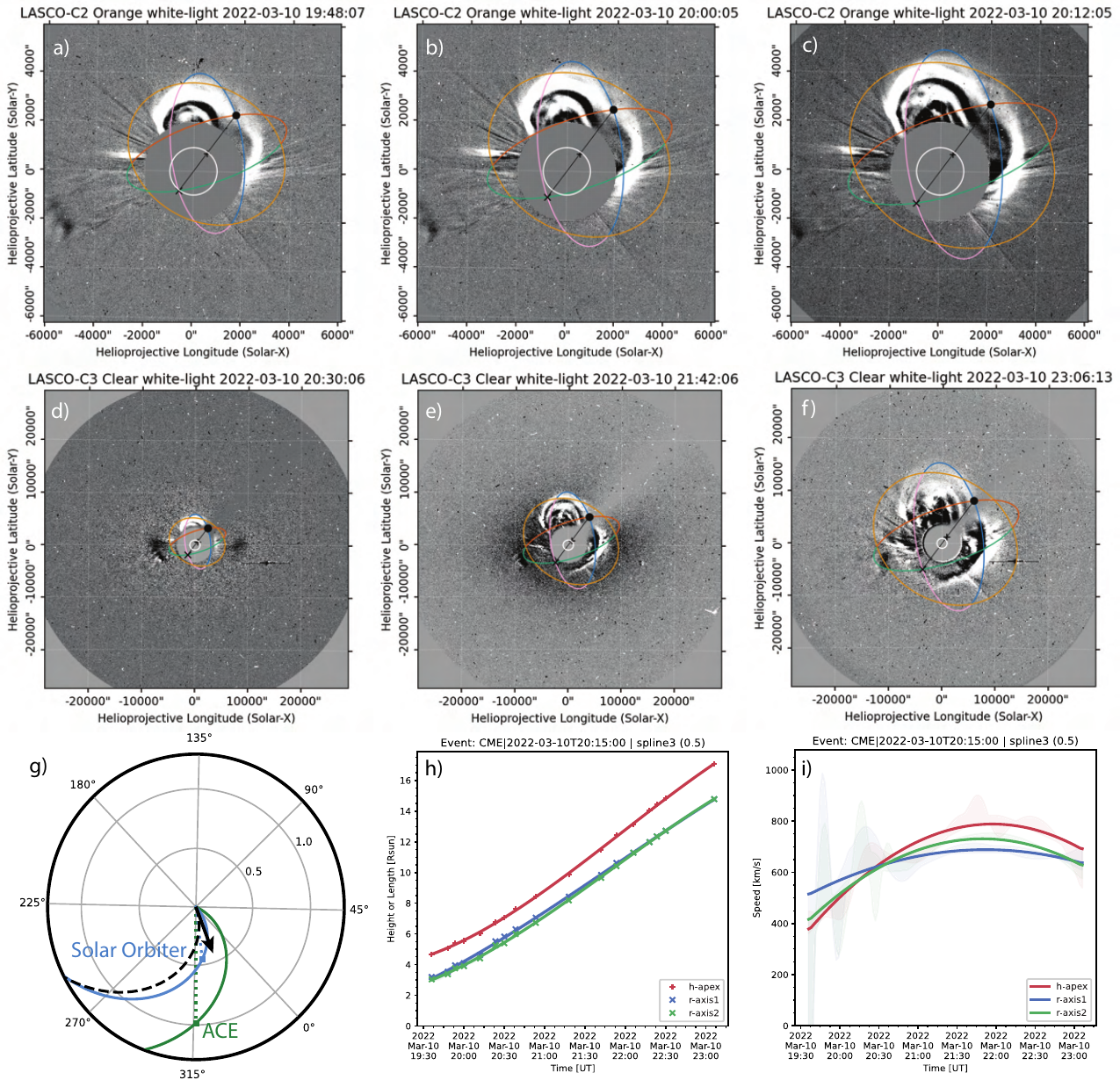}}
   \caption{Panels a-f show images from LASCO-C2 and LASCO-C3 depicting the evolution of the CME. These images also show the reconstructed shock wave calculated using an ellipsoidal model (orange lines) via the PyThea code package \citet{koul2022}. Panel g shows the relative positions of Solar Orbiter and ACE in Carrington coordinates during the initial stages of the event, shown in a view of the ecliptic plane from the ecliptic north, produced using the Solar-MACH tool \citep{gies2023}. The arrow of Panel g denotes the direction of the CME-driven shock nose expansion computed from PyThea. Panels h and i show the computed kinematic properties (height and speed) of the reconstructed shock wave from PyThea.}
              \label{Figcomp}%
    \end{figure*}
    
For this study, data from the Magnetometer \citep[MAG;][]{horb2020}, Solar Wind Analyser - Proton-Alpha Sensor \citep[SWA-PAS;][]{owen2020}  and the Energetic Particle Detector \citep[EPD;][]{rodr2020, wimmer2021} on Solar Orbiter \citep{muller2020} are utilized. EPD consists of the SupraThermal Electrons and Protons (STEP), the Electron Proton Telescope (EPT), the Suprathermal Ion Spectrograph (SIS), and the High-Energy Telescope (HET) sensors. For the Advanced Composition Explorer \citep[ACE;][]{stone1998} at 1 au, observations from the Magnetometer \citep[MAG;][]{smith1998}, Solar Wind Electron, Proton, and Alpha Monitor \citep[SWEPAM;][]{mc1998}, the Electron, Proton, and Alpha Monitor \citep[EPAM;][]{gold1998} and the Ultra Low Energy Isotope Spectrometer \citep[ULEIS;][]{mason1998} were utilized. These instruments provide magnetic field vectors (MAG from both spacecraft); thermal ion velocity, density, and temperature measurements (SWA-PAS and SWEPAM); and ion particle flux over a broad range of energies (2-80 keV for STEP, 25 keV - 6.4 MeV for EPT, and 6.8-107 MeV for HET on EPD at 0.45 au,  50 keV - 5 MeV for EPAM at 1 au). Measurements from these instruments were also used to compute various plasma parameters at the two heliospheric distances (see Table \ref{tabPlasma}, with methods given in Appendix \ref{app}).

For suprathermal ion composition, we also analyzed particle flux and energy data from SIS onboard Solar Orbiter and ACE ULEIS. Both instruments are high resolution time-of-flight (TOF) mass spectrometers that identify ion mass and energy through measurements of both TOF and residual kinetic energy of particles that enter the telescope acceptance cone. While SIS is well suited to measure ion species from H - Fe, ULEIS was optimized for species from He - Fe, with a very low efficiency (<1\%) for H particles. As such, this study focuses on \textsuperscript{4}He, C, O, Fe from both SIS and ULEIS as well as H from only SIS. 

To compute the initial kinematics of the CME-driven shock expansion near the Sun, this study uses the open-sourced PyThea code package \citep{koul2022} with remote sensing observations from the Large Angle Spectroscopic Coronagraph \citep[LASCO;][]{bru1995} onboard the Solar and Heliospheric Observatory spacecraft \citep[SOHO;][]{dom1995}, specifically its two externally occulted telescopes, LASCO-C2 and LASCO-C3, as well as the coronagraph COR2 onboard STEREO-A (not shown, \citet{howard2008}). Through fitting remote-sensing observations to an ellipsoid model, it is able to calculate the kinematics and three-dimensionally reconstruct shock waves.

This study also contains snapshots of simulation results from the online tool developed at the Community Coordinated Modeling Center (CCMC), to glean inner heliospheric conditions during the transit of the ICME. To reproduce these conditions, the tool utilizes the WSA-ENLIL+Cone Model, which is a 3D Magnetohydrodynamic modeling system that uses remote observations of both white-light signatures of CMEs in coronographs and the photospheric magnetic field to create a time-dependent description of the background solar wind plasma and magnetic field \citep{od2004,od2020}. CME-like hydrodynamic structures can then be inserted into this environment, bypassing a consideration of CME initiation, and the evolution of their kinematic properties simulated via ENLIL (for a detailed review, see \citet{mays2015}). The simulation includes CME and space weather data provided by the Space Weather Database of Notifications, Knowledge, Information (DONKI), and both the set-up parameters for the model and the full simulations results can be found on the CCMC website \footnote{\url{https://kauai.ccmc.gsfc.nasa.gov/DONKI/view/WSA-ENLIL/19410/1}} .

\subsection{Event Overview}

\begin{figure*}
   \centering
            {\includegraphics[width = \hsize]{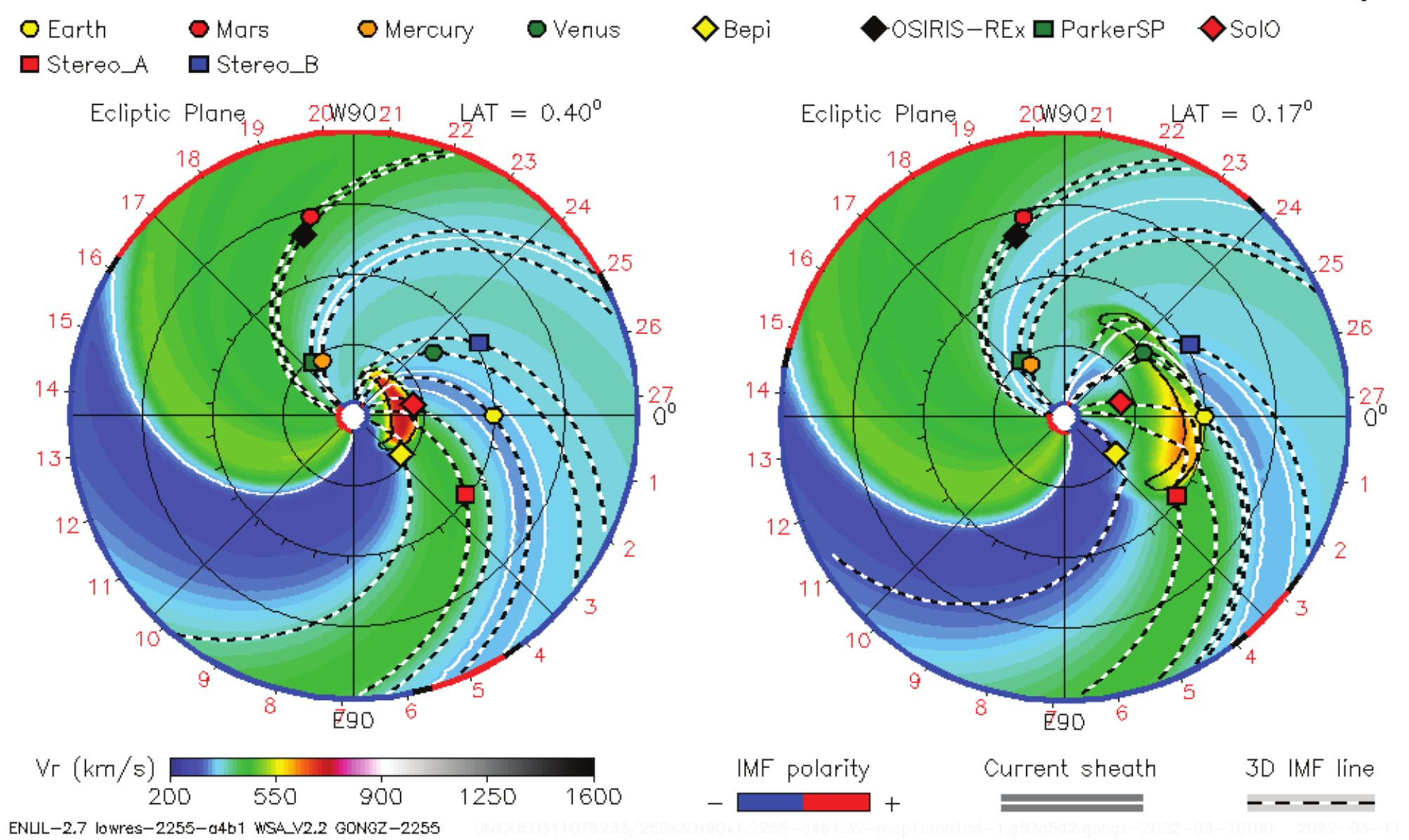}}
   \caption{Inner heliospheric radial solar wind velocity in the ecliptic plane simulated using the WSA-ENLIL + Cone Model provided by DONKI. The snapshots are taken at 2022 March 12 00:00 (left) and 2022 March 13 12:00 (right), corresponding to the time intervals of the ESP phase seen at Solar Orbiter and ACE. A SIR, denoted by the sharp transition between a slow solar wind stream (blue) and a faster stream (green), can be seen interacting with each spacecraft at the time of ICME passage. }
              \label{FigAG3}%
    \end{figure*}

\begin{figure*}[htbp]
   \centering
   \includegraphics[width = 16 cm, height = 16cm]{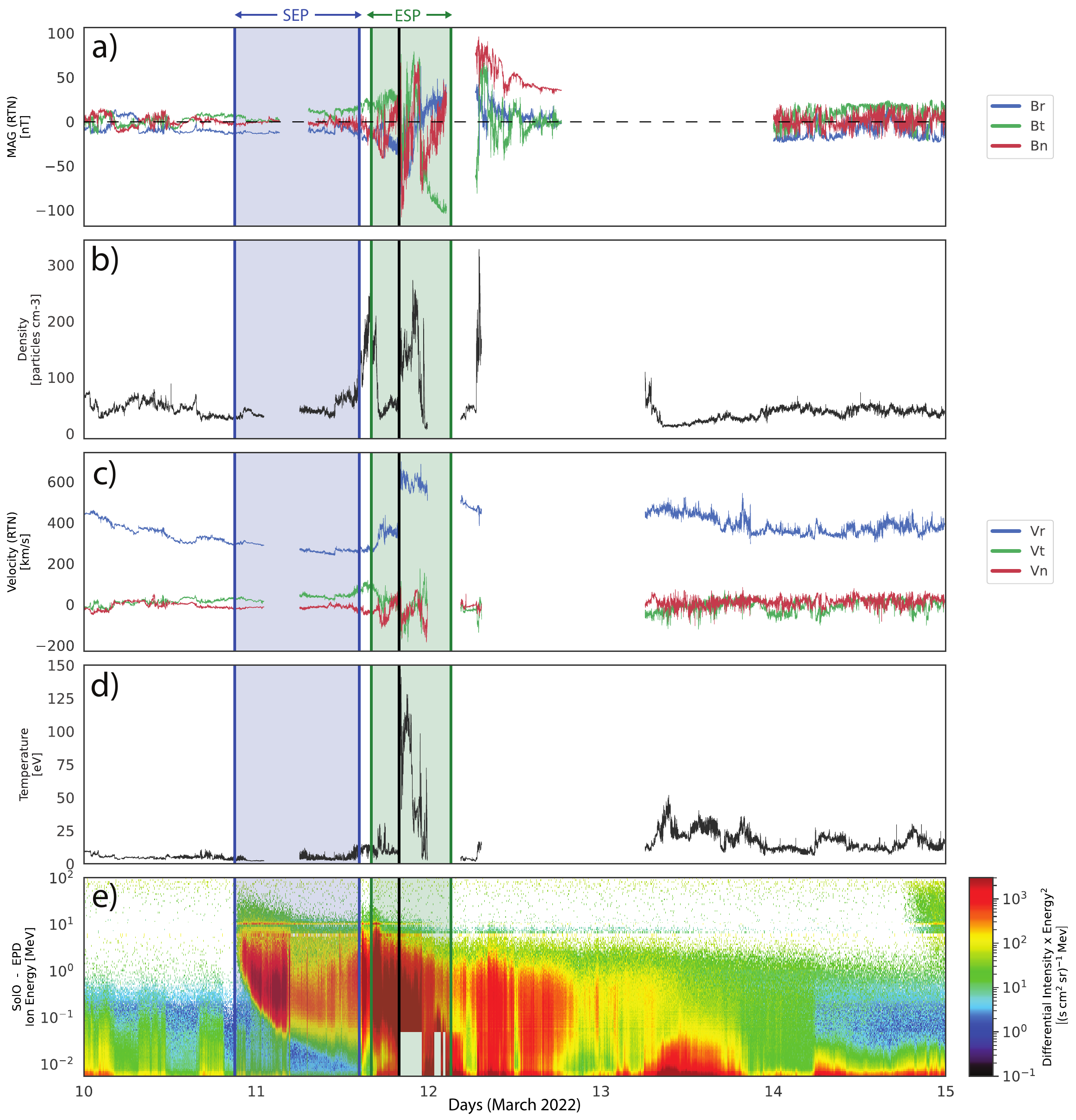}
   \caption{Overview of the Solar Orbiter measurements of the event. a) Magnetic field vector in RTN coordinates; solar wind ion (proton) b) density, c) velocity in RTN coordinate, and d) temperature; and e) color spectrogram of ion flux, with the color indicating the differential particle flux, scaled by the energy squared for better visibility. The blue and green lines enclose the gradual SEP (2022-03-10 21:00:02 to 2022-03-11 14:20:02) and ESP (2022-03-11 16:02:24 to 2022-03-12 03:07:24) phases of the event, respectively. The black line marks the passage of the shock jump at 2022-03-11 19:52:24. The decrease in ion intensities seen during the SEP time interval in panel e are likely due to an ion dropout, while the grey rectangles seen after the shock passage are the result of data gaps. }
              \label{FigSolo}%
    \end{figure*}

The eruption of the CME of interest occurred at around 17:10 UT on 2022 March 10 from active region 12962 (N27W07) with no accompanying major flare (for a review of the event, see \citet{zhuang2024} ). Figure \ref{Figcomp} illustrates a reconstruction of the CME-driven shock observed on 2022 March 10. This reconstruction was made using PyThea, which allows for the calculation of kinematics for the event via the spline fitting of the height-time measurements using the ellipsoid model. Panels a-c (d-f) show a subset of white-light images used to reconstruct the shock, specifically those taken by LASCO-C2 (LASCO-C3). The lines superimposed on these images denote the ellipsoid model fit for the reconstructed shock wave front. The time evolution of both the height and velocity of the shock are shown in panels h and i, respectively (for a detailed review of the PyThea code package, see \citet{koul2022}).  From this we determined that maximum speed of the shock at its apex was 788 km/s, and the ICME-driven shock nose expanded in the direction of $337^{\circ}$ longitude and $21^{\circ}$ latitude in Carrington coordinates.

Figure \ref{Figcomp} panel g shows a view of the spacecraft locations and associated Parker spirals (299 km/s for Solar Orbiter and 373 km/s for ACE) on the ecliptic plane relative to the Sun, along with the direction of ICME-driven shock nose expansion shown by the black arrow. Solar Orbiter and ACE have a longitudinal separation of  $7.2^{\circ}$ and a latitudinal separation of $2.9^{\circ}$  at this time, making the two spacecraft nearly radially aligned. Both spacecraft are also slightly offset from the direction of shock nose propagation, with Solar Orbiter being within $21^{\circ}$ longitude and $29^{\circ}$ latitude of the propagation direction, and ACE being within $14^{\circ}$ longitude and $26^{\circ}$ latitude of the propagation direction.

For a broader perspective of the state of the heliosphere during this event, Fig. ~\ref{FigAG3} provides the simulated inner heliospheric plasma conditions at the time of the ICME-shock arrival at Solar Orbiter (left) and ACE (right) using the WSA-ENLIL+Cone Model. While the simulation referenced was run using parameters compiled by outside sources rather than from those derived from Pythea, the CME speed input for the model ($677$ km/s) and the reconstructed ICME-associated shock speed do roughly agree, and both sets of reference longitude and latitudes are the same. Additionally, the model expects the arrival of the ICME shock at Solar Orbiter at around 2022-03-11 21:33, and at ACE at around 2022-03-13 11:35. These times also roughly agree with observations of the shock jump seen in the in-situ data. During these intervals, two SIRs likely exist in the heliosphere, both located at the sharp transition from slow solar wind speeds ($\sim 300$ km/s, blue) to faster speed streams ($\sim 450$ km/s, green). Coincidentally, one of these SIRs is suggested to intersect Solar Orbiter and ACE at the same time as the ICME arrival (Fig. ~\ref{FigAG3}, left and right panels, respectively). As such, the ICME-associated shock may have access to suprathermal seed populations with a composition typically observed at larger heliospheric distances, typical of SIRs. It is important to note that the in situ plasma data, while similar, does not exactly match the simulated data. Although robust modeling of this event is outside the scope of this paper, a more accurate reconstruction would be a worthwhile venture.

\begin{figure*}
   \centering
   \includegraphics[width = 17 cm, height = 16cm]{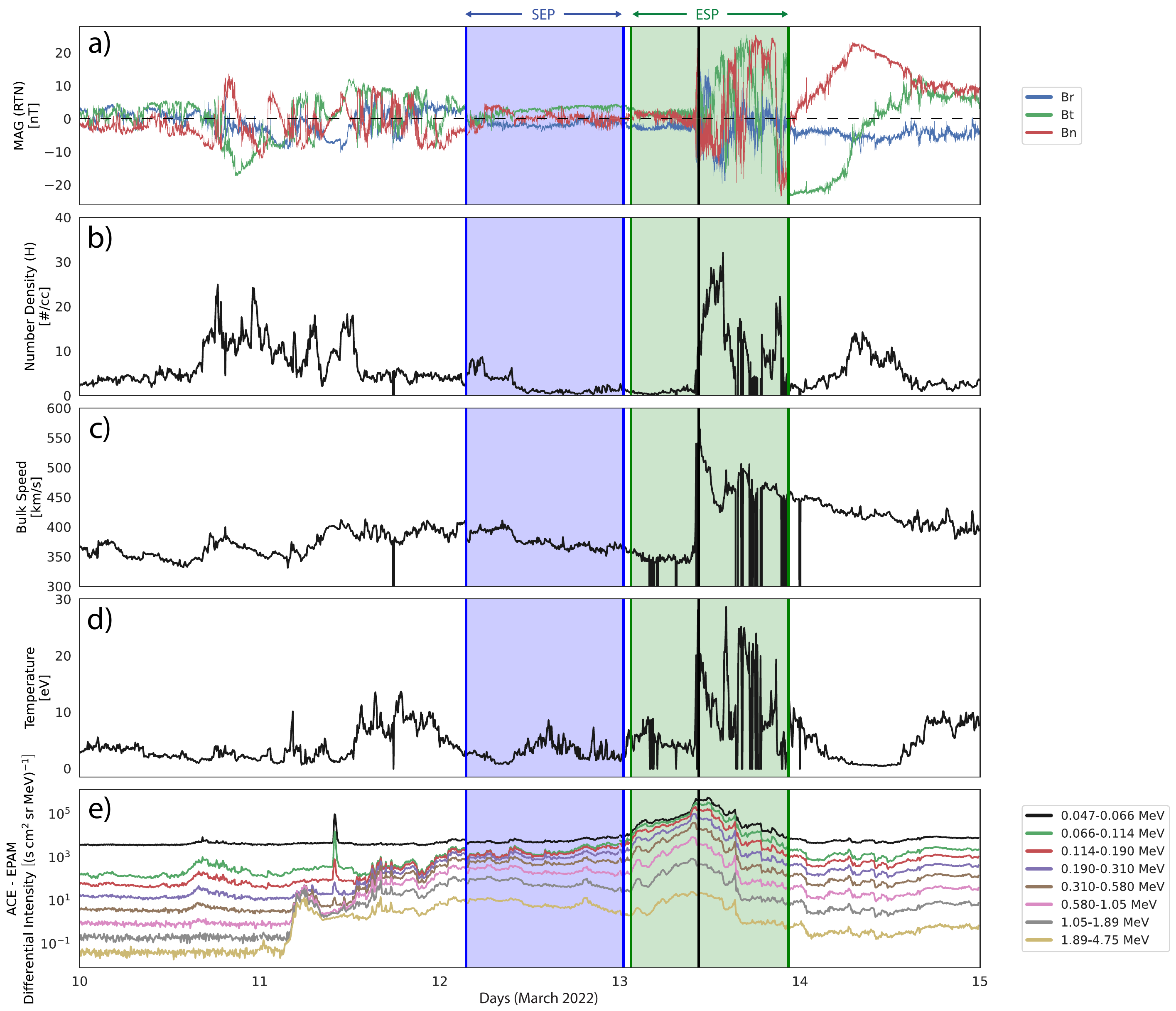}
   \caption{Overview of the ACE measurements of the event. a) Magnetic field vector with coordinates shown in the upper right corner; solar wind ion (proton) b) number density, c) bulk speed, and d) temperature; and e) ion flux, with the energy bins shown on the upper left corner. The blue, green, and black lines once again show the gradual SEP phase (2022-03-12 03:29:33 to 2022-03-13 00:30:33), ESP phase (2022-03-13 01:30:02 to 22:29:43), and passage of the shock jump (2022-03-13 10:29:51) respectively.}
              \label{FigACE}%
    \end{figure*}

Figures \ref{FigSolo} and \ref{FigACE} provide an overview of the ICME event at Solar Orbiter and ACE, respectively. The top panels on both figures show the components of the magnetic field in the Radial-Transverse-Normal (RTN) coordinate frame typically used to describe the orbit of a spacecraft, where $\hat{\mathrm{R}}$ is parallel with the radial vector, $\hat{\mathrm{N}}$ is parallel with the normal vector, and $\hat{\mathrm{T}}$ completes the orthonormal frame \citep{mac2014}. Panels b, c, and d show the ion density, velocity in the RTN frame, and temperature for Solar Orbiter (Fig. ~\ref{FigSolo}), with the blank sections being a result of data gaps, and the solar wind proton number density, bulk speed, and radial component of the proton temperature for ACE (Fig. ~\ref{FigACE}). Lastly, energetic particle flux is shown in Fig. \ref{FigSolo} and Fig. \ref{FigACE} panel e. For Solar Orbiter, ion flux measurements from STEP, EPT-Sun, and HET-Sun are combined into a single spectrogram and shown in units of flux times the square of the energy to better illustrate the full range of energies (Fig. ~\ref{FigSolo} panel e). Note that the sunward fields of view on both EPT and HET in Fig.  \ref{FigSolo} are chosen to match STEP, which only has a sunward-facing aperture, and to include particles streaming away from the Sun.   Meanwhile, ACE EPAM observations (Fig. ~\ref{FigACE} panel e) are represented as a time-flux line plot due to the limited number of energy channels. 

The blue lines in each panel denote the gradual SEP portion of the event. Identification of this interval was primarily driven by the observation of the dispersive, "1/V" signature seen in the ion flux due to time-of-flight effects of particles arriving from a remote source (\citet{mason2012,giac2020}, for review see \citet{desai2016}). As such, this interval begins with the arrival of the highest energy SEP onset particles observed at each spacecraft and ends when onset particles of the lowest observable energy appear, but before their significant increase in flux. An additional criterion for delineating the gradual SEP time interval was a steady interplanetary magnetic field, as to avoid flux enhancements potentially related to other dynamics, such as current sheet crossings or times when the spacecraft is within a flux rope structure.  The interval used for the gradual SEP portion of the event seen at Solar Orbiter is 2022-03-10 21:00:02 to 2022-03-11 14:20:02, while for ACE the interval is 2022-03-12 03:29:33 to 2022-03-13 00:30:33.  

Regarding the selection of the gradual SEP time interval seen by ACE, it is important to note that there is evidence that this portion of the event might have started earlier, specifically on 2022 March 11, where the flux of higher energy particles seen in Fig.  \ref{FigACE} panel e initially begin to increase. However, between this initial increase and beginning of the chosen gradual SEP interval, we observed a loss of \textsuperscript{3}He signatures, which we believe signifies a loss of connectivity to the active region from which the CME originated \citep{desai2006}. Additionally, the change in the sign of the radial magnetic field as seen by ACE during this time, signifying a change in source magnetic polarity, also suggests a loss of connectivity to the initial acceleration site of SEPs. We believe that this interval is instead representative, at least partially, of the SIR passage suggested by Fig. ~\ref{FigAG3}, as it is characterized by a decrease in particle density along with an increase in proton temperature. Both of these observations can be interpreted as signatures of an SIR, specifically the passage of the stream interface, or the boundary separating the fast and slow streams \citep{gosling1978, jian2006, dresing2023}. However, since CME-SIR interactions are difficult to fully discern in situ, especially in the case of their collocation at the observation site (e.g., \citet{burlaga2003}),  we concluded that these particles are not entirely SEP onset particles, and decided not to include them in our analysis nor designated this interval in Fig.  \ref{FigACE} . While this choice might impact the gradual SEP spectra detailed in Fig. ~\ref{FigResults}, it does not affect the conclusions of this study.

The ESP portion of the event was identified mainly by the additional particle enhancement surrounding the ICME driven shock (shown by the green region in each panel). During this time interval a gradual rise in the particle flux for all energy bins was identified, with the peak occurring at the passage of the shock itself, corresponding with sharp increases in the ion velocity and temperature, designated by the black line in Fig. ~\ref{FigSolo} and Fig. ~\ref{FigACE}. After this peak, the particle fluxes decrease, becoming constant following the passage of the ICME sheath. This combination in the observed particle flux of a gradual enhancement, peak at the shock front, and sharp decrease to a nearly constant downstream flux is normally seen in the ESP phase of an SEP event, and is well predicted by DSA theory \citep{lario2003,desai2004}. The interval used for the ESP portion of the event seen at Solar Orbiter is 2022-03-11 16:02:24 to 2022-03-12 03:07:24, with shock jump passage 2022-03-11 19:52:24, while for ACE the interval is 2022-03-13 01:30:02 to 22:29:43, with shock jump passage at 2022-03-13 10:29:51.

Lastly, we recognize that panel~e of Fig. ~\ref{FigSolo} exhibits some peculiarities, with the first being a wholesale decrease in differential intensity across energy channels during the latter half of the SEP time interval, and the second being the appearance of gray rectangles after the observed passage of the shock front. The decrease in intensity is likely the effect of an ion dropout caused by a change in magnetic connection to the source region on the shock front via interplanetary magnetic field (IMF) line meandering, as these have been known to occur for SEP events \citep{ho2022,giac2021}. The appearance of gray rectangles, on the other hand, are due to saturation effects in the STEP sensor, and consequently appear as data gaps. . 

\begin{table*}[t!]
    \centering
    \ra{1.5}
    \caption{Summary of plasma and shock parameters observed at the Solar Orbiter and ACE. Parameters calculated using methods from IPshocks, which are reviewed in Appendix \ref{app}.}
    \begin{tabular}{lcccccccccccc} \toprule
    &&&&&&&&&&\multicolumn{3}{c}{$ \mathbf{\hat{n}_{sh}}$}\\ \cmidrule(lr){11-13}
       & $\mathrm{r} \: (\mathrm{au})$& $C_{s}^{up} \: (\mathrm{km/s})$ & $V_{A}^{up}\: (\mathrm{km/s})$ & $V_{ms}^{up}\: (\mathrm{km/s})$ & $\beta^{up}$ & $M_{A}$ & $M_{ms}$ & $V_{sh}\: (\mathrm{km/s})$ & $\theta_{Bn}$ & $r$ & $t$ & $n$ \\ \midrule
       Solar Orbiter  & 0.45 &  73.2 & 150.1 & 167 & 0.3 & 7.2 & 6.5 & 742.8 & 29.2 & -0.98 & 0.21 & 0.01\\
       ACE  & 1 & 54.9 & 71.3 & 90 & 0.7 &  &  &  &52.3  & -0.70 & -0.33 & 0.63\\
    \end{tabular}
    \label{tabPlasma}
\end{table*}
 
\section{Results}\label{Results}
\begin{figure*}
   \centering
            {\includegraphics[width = 0.85\hsize]{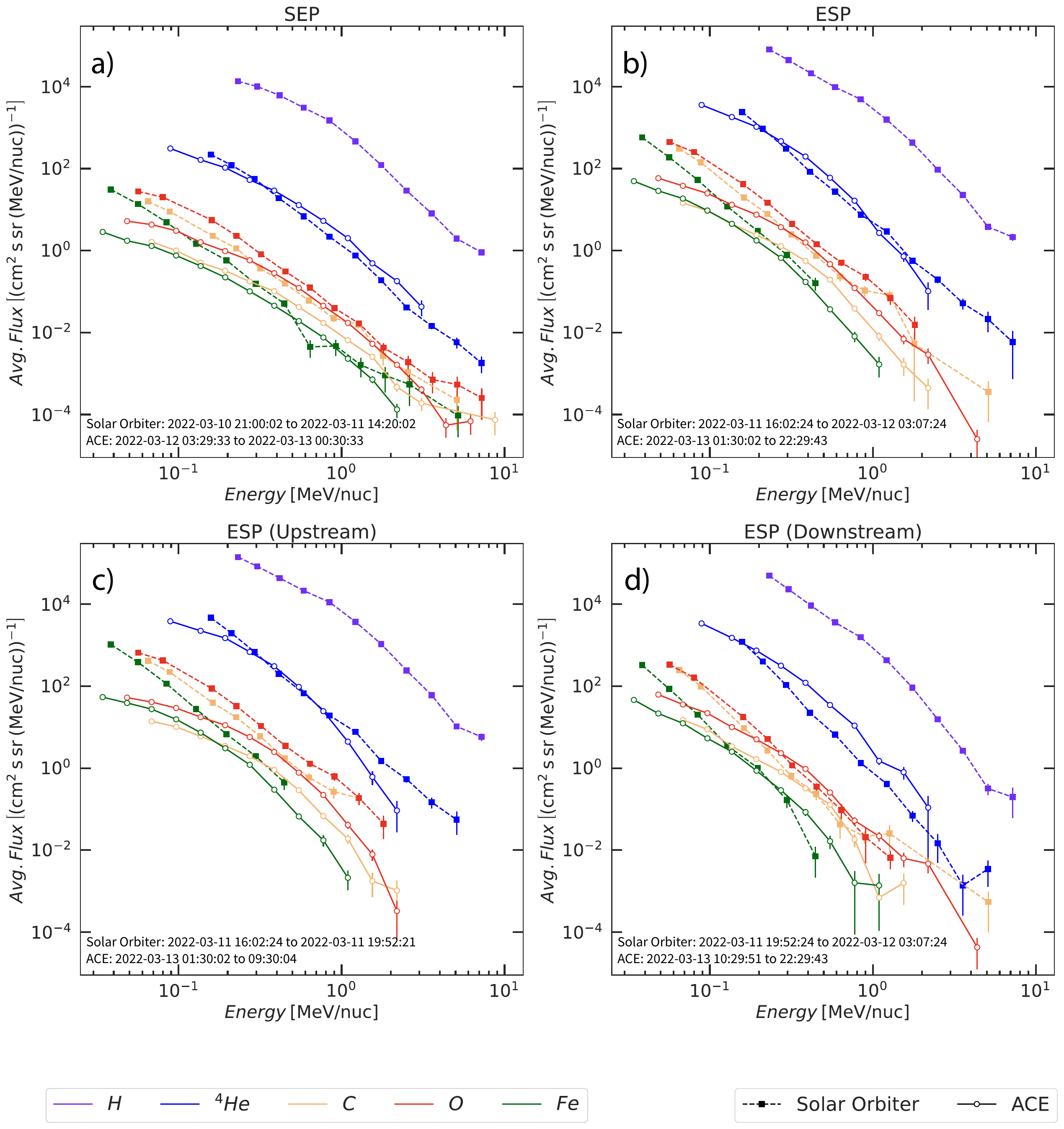}}
    
   \caption{Spectra for H, \textsuperscript{4}He, C, O, and Fe observed at Solar Orbiter (solid squares with dashed lines) and ACE (open circles with solid lines). The flux for these species are averaged over different time intervals: a) gradual SEP phase, consisting of particles accelerated close to the Sun, b) ESP phase, consisting of particles accelerated locally at the shock, and c and d separate the ESP phase into two portions consisting of particles that arrived before and after the shock jump, respectively. Measurements of H taken by ACE/ULEIS are excluded due to their low efficiency. The exact time intervals used are included on each panel. }
              \label{FigResults}%
    \end{figure*}

Figure~\ref{FigResults} shows the time-averaged particle spectra for H, \textsuperscript{4}He, C, O, and Fe ions taken between the gradual SEP and ESP intervals listed in the caption of Fig. ~\ref{FigSolo} and Fig. ~\ref{FigACE} and in the text above. The solid and dotted lines denote the spectra taken at Solar Orbiter and ACE, respectively, while the colors of each line denote the ion species. These spectra were compiled by averaging the flux over the time intervals corresponding to the arrival of the gradual SEP and the ESP phases (panels~a and~b) for each energy bin within the spectrograms of both spacecraft that had reliable counting statistics. As noted in Section~ \ref{inst}, H fluxes are only included for Solar Orbiter. 

Additionally, Fig. ~\ref{FigResults}~panels~c and~d show the ESP spectra upstream and downstream of the shock. This was done to separate the different processes that shape particle spectra, such as particle escape in the upstream plasma (e.g., \citet{sch2015}, \citet{desai2016b}) and particle trapping in the downstream plasma (e.g., \citet{zank2015}, \citet{leroux2015}). These processes have the capacity to alter the measured particle spectra, resulting in different spectral slopes in different regions of the shock-associated plasma.

To analyze the shape of the measured particle spectra, the spectra are fitted either using a single power law or using the pan spectrum fitting formula from \citet{liu2020}\begin{equation}
J(E)=A \times E^{-\beta_1}\left[1+\left(\frac{E}{E_0}\right)^\alpha\right]^{\frac{\beta_1-\beta_2}{\alpha}},
\end{equation}where $J$ is the particle flux, $E$ is the particle energy, $A$ is the amplitude coefficient, $E_0$ is the spectral transition energy, $\alpha$ represents the sharpness and width of the spectral transition region, which is centered around $E_0$, and $\beta_1$ and $\beta_2$ are power law indices that describe the spectral shape before and after this spectral transition region, respectively. For our study, both $E$ and $E_0$ are in units of kinetic energy per nucleon. This formula was specifically developed to describe the representative energy spectra from various suprathermal particle phenomena, including gradual SEPs and ESPs. The results of this fitting can be seen in table ~\ref{tabFit}.

From a combination of visual analysis and spectral fitting, we find that for each time interval, the spectra measured at ACE tend to have spectral breaks, exhibiting a shape more accurately described by an ER function.  On the other hand, the spectral shapes for the ion species measured at Solar Orbiter do not have these breaks, and are better fit by a single power law. In general, the spectra measured at Solar Orbiter are much steeper (i.e., larger negative slope) when compared to those taken at ACE for energies below the spectral break in the ACE spectra. However, the spectral slopes observed at ACE for energies above the spectral break closely match those observed at Solar Orbiter for all energies measured by SIS. Additionally, the average flux at these higher energies are similar at both spacecraft, rather than being lower at ACE due to volumetric expansion.

\section{Discussion}\label{Disc}
\begin{figure*}
   \centering
            {\includegraphics[width = \hsize]{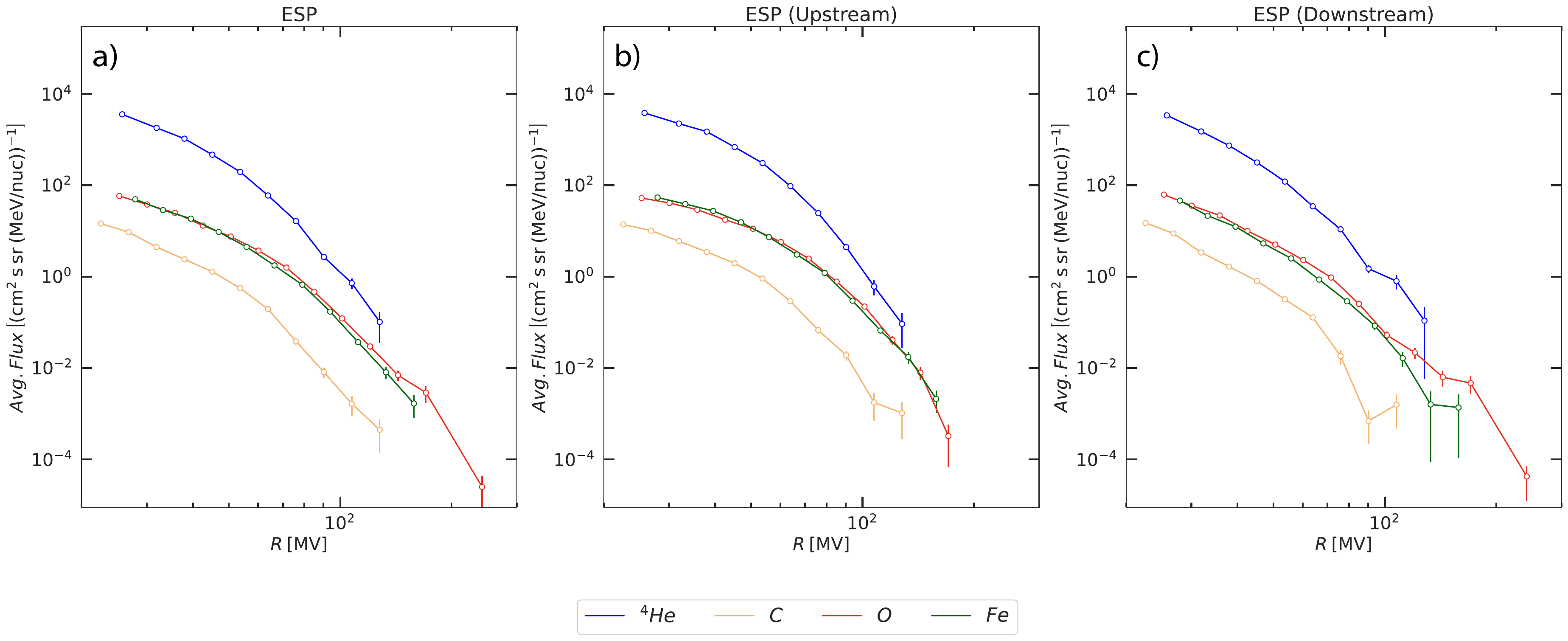}}
    
   \caption{The rigidity-scaled spectra for the species observed at ACE. The time intervals used in panels a-c are the same as panels b-d in Fig. \ref{FigResults}.  }
              \label{FigAG1}%
    \end{figure*}

\begin{figure*}
   \centering
            {\includegraphics[width = \hsize]{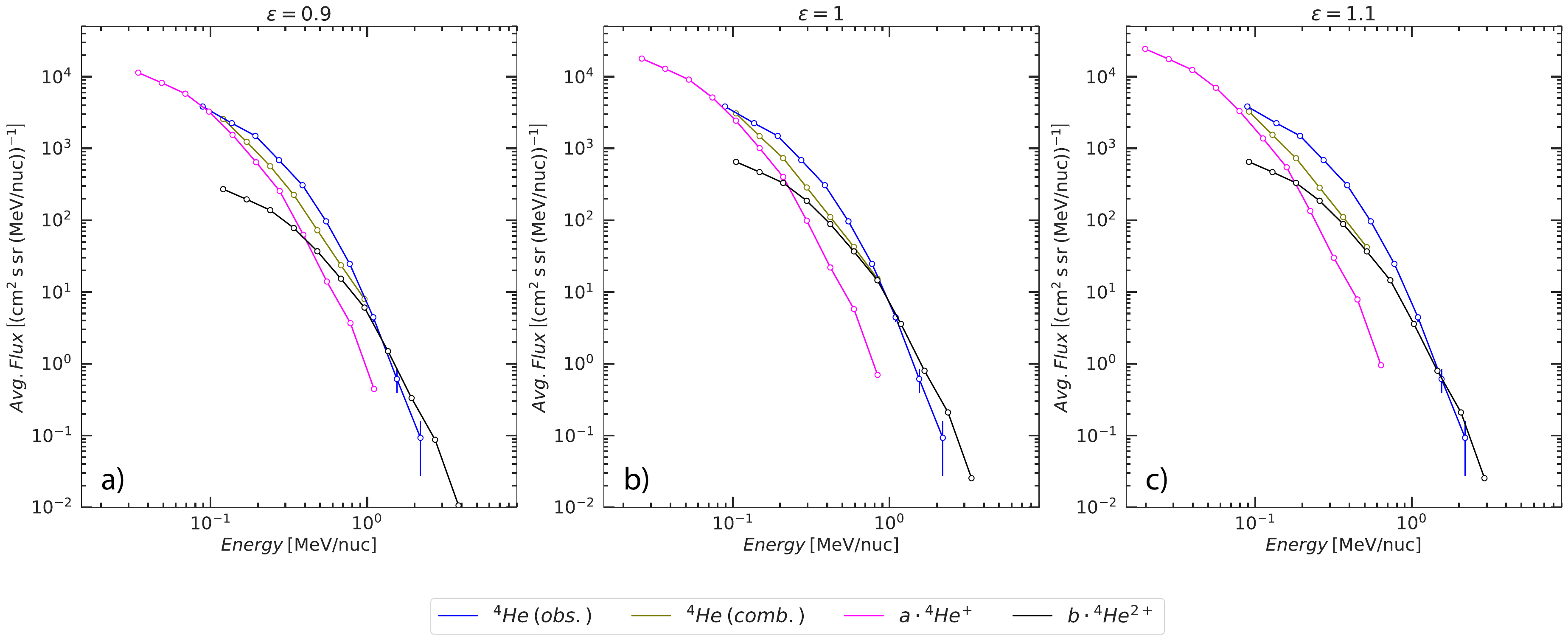}}
     
   \caption{Reconstruction of the \textsuperscript{4}He spectra observed at ACE in the upstream ESP time interval. Consists of three portions: the pickup ion ($\mathrm{He^{+}}$, magenta) and alpha particle ($\mathrm{He^{2+}}$, black) spectra, made by shifting the rigidity-scaled Fe spectra into energy space using equation ~\eqref{rigidity}, albeit with the appropriate charge and mass values, and the combination of the two, denoted by the olive line the area where the individual spectra overlap. Panels a - c show the reconstruction using different values of $\epsilon$ from equation \eqref{kappa}. Constants $a$ and $b$ are used to alter the relative concentrations of the two particle populations.}
              \label{FigAG2}%
    \end{figure*}

    
Differences in the gradual SEP spectra at each spacecraft (Fig. ~\ref{FigResults} panel~a), while not the focus of this paper, can likely be explained as a result of the spacecraft being magnetically connected to different portions of the shock lower in the corona, as it has been shown that certain areas along the shock are prone to accelerating particles to higher energies (e.g., \citet{chen2022,gopal2014,gopal2016,gopal2018}) in addition to the ion dropout in the lower energy range. However, the observation of similar spectral differences during the ESP phase of the event warrants further analysis. We would expect from their radial alignment that the spacecraft would measure particle spectra with similar slopes, albeit with a decrease in average flux across all energies at 1 au due to volumetric expansion, as both spacecraft survey the same portion of the shock structure. The fact that spectral breaks are only observed at ACE with the high energy portion having nearly the same spectral index and average flux at both spacecraft suggests continued evolution of the particle populations between 0.5 and 1 au and/or varied local shock geometries at each spacecraft.

Both spectral breaks and higher energy power laws in double-power law spectra associated with ESP events are highly dependent on rigidity. This connection is primarily due to the dependence of the diffusion coefficient, which influences particle diffusion from the shock region, on the charge-to-mass ratio. This ratio is in turn related to magnetic rigidity, or the momentum per unit charge, as it is defined in our study as in \citet{reames2020}\begin{equation}\label{rigidity}
    P=\frac{pc}{\mathrm{Q}e} = \frac{M_{u} \beta \gamma \mathrm{A}}{\mathrm{Q}},
\end{equation} where $\beta = v/c$ is the particle velocity relative to the speed of light, $\gamma = (1-\beta^{2})^{-1/2}$, $\mathrm{Q}$ is the ion charge in the unit of $e$, $\mathrm{A}$ is the atomic mass,  $M_{u} = m_{u}c^{2} = 931.494 \: \mathrm{MeV}$, which is related to energy (in $\mathrm{MeV \: amu^{-1}}$) via the equation \citep{reames2020}\begin{equation}\label{rigidity2}
    E = \mathcal{E} / A = M_{u}(\gamma - 1) \approx \frac{1}{2} M_{u} \beta^{2},
\end{equation} where $\mathcal{E}$ is the total kinetic energy. In the general case, the diffusion coefficient can be defined as \citep{li2009}\begin{equation}\label{kappa}
\kappa=\kappa_0\left(v / v_0\right)^\tau(\mathrm{A / Q})^\epsilon,
\end{equation} where $v_0$ is the speed of particle injection into the shock structure, and $\kappa_0$ is the reference diffusion coefficient at $v_0$. $\tau$ and $\epsilon$ are both constants. Equation~\eqref{kappa} clearly displays the correlation between the diffusion coefficient and the charge-to-mass ratio, $\mathrm{Q/A}$. The relationship between $\kappa$ and the spectral shape of gradual SEPs stems from the equal diffusion coefficient condition, which determines the location of the spectral break for heavy ion spectra. This condition states that the break energies of heavy ions in an event occur at the same $\kappa$ value. Since particle escape from the shock region is regulated by $\kappa$, the specific $\mathrm{Q/A}$ dependence found in this coefficient will carry over to the break energy $E_0$. From this condition, the dependence of the break energy on $\mathrm{Q/A}$ can be defined as \citep{li2009} \begin{equation}\label{break}
E_0=\tilde{p}_m^2 \sim\left(\frac{\mathrm{Q}}{\mathrm{A}}\right)^{2 \epsilon / \tau} ,
\end{equation} where $p_m$ is the maximum particle momentum per nucleon. 

Such a correlation has been observed in a number of ESP events, typically in the form of $\mathrm{Q/A}$ ordering of heavy ion spectra fit with double power laws or exponential rollover models. In these studies, the ability to order the break energies of these spectra using $(\mathrm{Q/A})^{\delta}$, where $\delta$ varied between events, typically within the range of $\approx 0.7-2$, suggests a difference in the underlying acceleration processes present in each (e.g., \citet{ty2000}, \citet{cohen2003,cohen2005}, \citet{mewaldt2005a,mewaldt2005b}).Therefore, scaling the particle spectra measured at ACE to rigidity would provide insight to whether the particle transport is diffusion dependent, and thus a product of shock acceleration, via an observed $\mathrm{Q/A}$ dependence of the spectral breaks.

Figure \ref{FigAG1} shows the ACE spectra in rigidity space assuming charge states of 2, 6, 6, and 16 for He, C, O, and Fe, respectively, with panel a displaying the spectra integrated over the entire ESP phase, and panels b and c showing the spectra for only the up- and downstream portions of the ESP enhancement. These charge states were chosen from the average SEP ionic charge states determined by \citet{mobius2000} and \citet{klec2007a}, which are $\mathrm{He}^{2+}$, $\mathrm{C}^{5.6+}$, $\mathrm{O}^{6.8+}$, and $\mathrm{Fe}^{11.6+}$ \citep{desai2016b}. Modeled ICME charge state compositions found in \citet{rakowski2007} and \citet{lam2023}, compiled using in situ data from various spacecraft at for 1 au, inform the exact chosen state from the cited average values. In each panel, the shapes of the heavier ion spectra exhibit noticeable similarities. The O and Fe spectra are nearly equivalent, and the C spectra has a similar shape, albeit shifted slightly to the left.  This suggests that their spectra are in accordance with DSA theory rather than an alternative acceleration mechanism. 

We note that the choice of $\mathrm{Fe}^{16+}$ for our analysis is a departure from both the value stated in \citet{desai2016b}, as well as the typical solar wind values for Fe charge states ($Q_{\mathrm{Fe}} \sim 9-11$, the most abundant being $10+$), which are generally assumed for gradual SEP events \citep{ko1999}. However, ICMEs have been shown to contain different charge state distributions when compared to solar wind values, especially for Fe. \citet{klec2007a} showed that while at low energies ($\sim$0.1 MeV/nuc) Fe mean charge states are similar to their solar wind values, at higher energies these charge states exhibit event-dependent variations, with  $Q_{\mathrm{Fe}} \sim 15-20$ at energies above $\sim$10 MeV/nuc. Similarly, \citet{gru2011} found that $95\%$ of ICMEs had Fe charge distributions that exhibited bimodality, with peaks at both $10+$ and $16+$. The significant abundance of $\mathrm{Fe}^{16+}$ reflected in both observations and in models, in addition to the fact that this charge state provides the best alignment with the spectra of C and O in rigidity space (not shown) ultimately solidified our choice.

On the other hand, the shape of the $\mathrm{^4He}$ spectra tends to be fundamentally different than that of the heavier ions, specifically in the upstream ESP region (Fig. \ref{FigAG1} panel b). The upstream ESP $\mathrm{^4He}$ spectra is characteristically different, in that matching the spectra cannot be achieved through a multiplicative factor on the flux value. Instead the spectral characteristics for $\mathrm{^4He}$ appear offset in rigidity space. This implies that, especially for these upstream particles, there is either preferential acceleration of this species outside of what is normally expected from shock acceleration or that the $\mathrm{^4He}$ population is comprised of a larger abundance of singly ionized pick-up ions than expected at 1 au (as a He population consisting of solely alpha particles, or doubly ionized helium, was assumed when calculating rigidity for Fig. \ref{FigAG1}).

A simple method for estimating the possible contribution of an overabundance of pickup ions at 1 au in the misalignment of the upstream ESP \textsuperscript{4}He spectra consists of finding a combination of alpha particles and pickup helium that could actually align with the other ion spectra in rigidity space. Starting with the rigidity scaled spectra for Fe, as it aligns well with the other spectra, representative spectra are made for He\textsuperscript{+} and He\textsuperscript{2+}. Then, with this basis we approximate both an alpha particle ($\mathrm{He^{2+}}$) and pickup helium ($\mathrm{He^{+}}$) spectra in energy space using the mass and charge for each, employing the formula for rigidity (equation~\ref{rigidity}), and solving for the energy spectra. Finally, we linearly combine these spectra and compare this reconstructed $\mathrm{^{4}He}$ spectra to the upstream observations. Figure \ref{FigAG2} shows the results of this effort, with the magenta and black lines showing the reconstructed pickup ion and alpha particle spectra respectively, and the olive lines showing the combination of the two. The blue lines denote the upstream ESP \textsuperscript{4}He spectra observed at ACE. Panels a - c show this process done while varying the scaling factor $\epsilon$ in Equation~\eqref{break}(with $\tau = 2$ implied). Note that the case shown in panel b where $\epsilon = 1$ correlates to the Bohm approximation. This approximation signifies that the scattering of the particles within the shock region occurs about once per gyroperiod, or the time it takes for a particle to complete a revolution around a magnetic field line, and has been found to work well as a heuristic in the regime of strong turbulence, where the ratio of the magnetic fluctuation, $\delta B$, to the background magnetic field energy, $B_0$, is greater than unity \citep{riordan2019}. From Fig. ~\ref{FigAG2} we find that the reconstructed, composite spectra consisting of both $\mathrm{He^{+}}$ and $\mathrm{He^{2+}}$ populations are able to roughly approximate the observed $\mathrm{^{4}He}$ spectra in panels a and b. This implies that pickup helium constitutes an essential portion of the observed $\mathrm{^{4}He}$ for these cases, as alpha particles are not able to reproduce the observed spectra alone. The remaining differences between the two are possibly a result of the actual pickup ion distribution having a fundamentally different spectral shape compared to that of both the heavier ions and the alpha particles for this event.

The high concentration of $\mathrm{He^{+}}$ relative to $\mathrm{He^{2+}}$ at 1 au, implied by the reconstructed spectra in Fig. \ref{FigAG2}  ,suggests the existence of a suprathermal pool abundant in $\mathrm{He^{+}}$ that is subsequently accelerated at the ICME shock. As shown in Fig. ~\ref{FigAG3}, the WSA-ENLIL+Cone Model model shows that a SIR may have been passing by both Solar Orbiter and ACE when the ICME would have been observed. SIRs typically form shocks further out in the Heliosphere \citep[e.g.;][]{jian2006,jian2008,pizzo1978,smith1976}, and the number density of pickup ions follows a  $r^-{2}$ dependence that peaks at 12 au \citep{zank2018, holzer1972, isen1986, khab1996, zank1999}.  A suprathermal pool of particles from a SIR could thus explain the abundance of $\mathrm{He^{+}}$ suggested at 1 au, as it may have a composition more reminiscent of the plasma environment beyond 1 au where pickup ions make up a significantly larger portion of the overall He population. There is evidence, both observational and computational, that suggests that SIRs can result in such a population of accelerated ions. For instance, \citet{mason2008} surveyed the heavy ion spectra and composition of He-Fe for 41 SIRs during 1998-2007 using ACE and found that SIR energetic particles are accelerated out of a suprathermal ion pool that includes heated solar wind ions, remnant suprathermals from impulsive SEPs, and pickup ions. \citet{morris2001} observed an increase in the abundance of $\mathrm{He^{+}}$ with time in SIRs at 1 au, a supposed byproduct of an improving magnetic connection between 1 au and efficient portions of the distant shock over the course of an event. Additionally, simulations by \citet{wij2023} found that the interaction between slow and high speed streams can significantly impact the suprathermal seed population in the inner heliosphere.

It is important to note that in Fig. ~\ref{FigAG2}, we assume that the ratio of $\mathrm{He^{+}}$ to $\mathrm{He^{2+}}$ at 1 au can approach 1 at a certain energy, which is around 0.35, 0.25, and 0.2 MeV/n for panels a through c, respectively. Below these energies, the $\mathrm{He^{+}}$/$\mathrm{He^{2+}}$ abundance ratio is dominated by $\mathrm{He^{+}}$ to varying degrees depending on the value of $\epsilon$. The ratio at 0.1 MeV/n is $\approx 8.5$ for $\epsilon = 0.9$, $\approx 3.7$ for $\epsilon = 1$, and $\approx 2.3$ for $\epsilon = 1.1$. Such high concentrations of pickup ions are unlikely at this heliospheric distance, as measurements of the abundance ratio within SIRs at energies $\leq 300$ keV and at 1 au have cited values of around 0.15 \citep{cho2000,klec2007b}. This suggests that matching the reconstructed He spectra to observations would require an excess of alpha particles at these energies, as there is still a significant gap between the two spectra even with the addition of an unrealistic concentration of $\mathrm{He^{+}}$. Furthermore, it suggests that the He spectrum cannot be directly scaled to the spectra of heavier ions in rigidity space for this event, as both the spectra of the pickup ions (as implied earlier) and the alpha particles have fundamentally different particle distributions than those of the heavier ions. 

The He ratios for the higher energies within the overlap between the $\mathrm{He^{+}}$ and $\mathrm{He^{2+}}$, however, are more reasonable. \citet{kuch2003} conducted a systematic study of the He ratio in the energetic population over the energy range of 0.25-0.8 MeV/n at 1 au and found that, while the long term average of the ratio was about 0.06, there was significant small scale variability caused by intermittent spikes in the abundance of $\mathrm{He^{+}}$ relative to $\mathrm{He^{2+}}$. Many of these $\mathrm{He^{+}}$ enhancements were attributed to interplanetary shocks and SIRs, with 33 \% of the observed 91 events showing abundance ratios exceeding 0.2, and 19\% exceeding 0.3. In this study, the large abundance ratios were mostly produced by SIRs, with the highest ratio value of 0.8 being the result of a series of recurring SIR passages. Motivated by \citeauthor{kuch2003} we compute the He ratio over the integrated energy range of 0.25 - 0.8 MeV/n for the spectra with $\epsilon$ values of 0.9 and 1. The spectra made using an $\epsilon$ value of 1.1 did not have an overlapping region that spanned the chosen energy range, so a similar analysis was not done. The abundance ratio calculated for the $\epsilon = 0.9$ case was 1.43, and the ratio for the $\epsilon = 1$ case was 0.43. These ratios for 0.25 - 0.8 MeV/n He are more realistic than those found at lower energies, with the latter being comparable to the helium abundance ratios observed in SIRs seen by \citeauthor{kuch2003}. The fact that the combined He spectra in this energy range continues to be closer to the observations than the $\mathrm{He^{2+}}$ spectra alone further suggests that the addition of pickup ions is required to describe the spectral shape of He in this event.   

Although the above analysis of rigidity effects at ACE suggests the existence of a suprathermal pool rich in pickup ions, this does not fully account for the differences between the spectra observed at 1 au and 0.45 au, namely the lack of spectral breaks in the spectra measured by Solar Orbiter. The cause of this discrepancy could lie in the effects of shock geometry. Through the use of nonlinear guiding center theory, \citet{li2009} found that the $\mathrm{Q/A}$ scaling parameter, $\delta$, differed significantly between quasi-perpendicular and quasi-parallel shocks reaching 1/5 and as high as 2, respectively. \citet{zhao2016} analyzed the spectra taken by three spacecraft, STEREO A (STA), STEREO B (STB), and ACE, during the 2013 November 4 gradual SEP event. The spacecraft were all stationed at 1 au, but were separated longitudinally, with the angle between ACE and STA being 148.56 degrees, and the angle between ACE and STB being 143.24 degrees. They found that STA and ACE had similar scaling parameters (1.12 and 1.01) while the scaling parameter measured at STB was significantly different (0.66). Using the results of \citet{li2009} as a framework, they were able to conclude that STA and ACE were connected to the quasi parallel portion of the shock, whereas STB was connected to the quasi perpendicular portion. This difference in surveyed shock geometry also affected the spectral shapes observed at each spacecraft: despite the higher energy portion of the spectra seen at STA being relatively steeper, spectral breaks can be seen at both STA and ACE, yet are practically non-existent at STB for the measured energy range. Surprisingly, despite being near radially aligned, the shock thetas for our event (Table \ref{tabPlasma}) suggest a similar difference in shock geometry between the two spacecraft, with Solar Orbiter connected to the quasi-parallel portion of the shock, and ACE to the quasi-perpendicular portion. Given the differences seen in \citet{zhao2016}, it is plausible that a similar effect is observed for our event, resulting in the breaks present at 0.45 to be out of the energy range observable by Solar Orbiter.

\section{Conclusions}\label{Conc}
\begin{sidewaystable*}[h]
    \centering
    \ra{1.4}
    \caption{Fitted parameters of Solar Orbiter and ACE spectra shown in Fig. ~\ref{FigResults}. Entries where only $\beta_1$ is shown represent spectra that are better fit by a power law (i.e $J(E)=A \times E^{-\beta_1}$).}
    \begin{tabular}{lccccccccc} \toprule
       & \multicolumn{4}{c}{$ \mathrm{H}$}& \multicolumn{4}{c}{$ \mathrm{^4He}$} \\
       Time Interval&$\beta_{1}$& $\beta_{2}$ & $E_0 \: \left(\mathrm{MeV}\right)$ & $\alpha$&  $\beta_{1}$& $\beta_{2}$ & $E_0 \: \left(\mathrm{MeV}\right)$ & $\alpha$\\ \midrule
       
       Solar Orbiter (SEP)  & $1.37 \cdot 10^{-9} \pm 1.98 $  &  $5.85 \pm 6.36$  & $1.02 \pm 1.37$ & $1.02 \pm 1.65$& 
       $3.14 \pm 0.04$ &  $\cdots$ &$\cdots$&  $\cdots$\\ 
       
       ACE (SEP)  &$\cdots$  &$\cdots$  & $\cdots$ & $\cdots$& 
       $1.35 \pm 0.28$ &$4.53 \pm 2.71$&$0.85 \pm 0.85$& $1.64 \pm 1.21$  \\ 
       
       Solar Orbiter (ESP)  & $2.25 \pm 0.03$ &  $3.93 \pm 0.49$ & $1.10 \pm 0.21$ & $15 \pm 15.86$& 
       $3.42 \pm 0.03$& $\cdots$ & $\cdots$ & $\cdots$\\ 
       
       ACE (ESP)  & $\cdots$ & $\cdots$ & $\cdots$ & $\cdots$& 
       $1.28 \pm 0.39$&$6.19 \pm 2.65$ & $0.58 \pm 0.33$& $1.72 \pm 0.92$ \\ 
       
       Solar Orbiter (ESP, up)  & $2.00 \pm 0.03$ &  $3.81 \pm 0.26$ & $1.04 \pm 0.08$ & $15 \pm 1.38$& 
       $3.24 \pm 0.04$&$\cdots$ &$\cdots$ &$\cdots$  \\ 
       
       ACE (ESP, up)  & $\cdots$ & $\cdots$ & $\cdots$ & $\cdots$& 
       $0.62 \pm 0.85$&$8.49 \pm 7.69$ &$0.81 \pm 1.08$&$1.29 \pm 0.98$  \\ 
       
       Solar Orbiter (ESP, down)  & $2.79 \pm 0.04$ &  $4.86 \pm 1.12$ & $1.27 \pm 0.35$ & $11.81 \pm 22.55$ & 
       $4.03 \pm 0.05$ & $\cdots$ & $\cdots$ & $\cdots$ \\ 
       
       ACE (ESP, down)  & $\cdots$ & $\cdots$ & $\cdots$ & $\cdots$& 
       $1.68 \pm 0.41$ &  $5.21 \pm 1.97$ &$0.47 \pm 0.26$&  $1.89 \pm 1.28$ \\ 
       \\
       
       & \multicolumn{4}{c}{$ \mathrm{C}$}& \multicolumn{4}{c}{$ \mathrm{O}$}& \\ \cmidrule(lr){2-5} \cmidrule(lr){6-9}\\

       Solar Orbiter (SEP)  & $ 2.42 \pm 0.06 $  &  $\cdots$ & $\cdots$ & $\cdots$& 
       $2.82 \pm 0.03$ &  $\cdots$ &$\cdots$&  $\cdots$\\ 
       
       ACE (SEP)  & $1.53 \pm 0.11$ &  $3.18 \pm 0.30$ & $0.50 \pm 0.08$ & $4.11 \pm 2.60$ & 
       $1.04 \pm 0.23$ & $4.01 \pm 0.69$ &$0.54 \pm 0.16$& $1.53 \pm 0.59$  \\ 
       
       Solar Orbiter (ESP)  & $3.05 \pm 0.06$ &  $\cdots$ & $\cdots$ & $\cdots$& 
       $3.17 \pm 0.05$& $\cdots$ & $\cdots$ & $\cdots$\\ 
       
       ACE (ESP)  & $1.77 \pm 0.06$ &  $4.76 \pm 0.42$ & $0.46 \pm 0.05\: \mathrm{MeV}$ & $4.95 \pm 1.63$& 
       $1.39 \pm 0.09$ & $4.78 \pm 0.35$ & $0.42 \pm 0.04$& $2.57 \pm 0.50$ \\ 
       
       Solar Orbiter (ESP, up)  & $2.37 \pm 0.07$ &  $\cdots$ & $\cdots$ & $\cdots$& 
       $3.02 \pm 0.05$& $ \cdots$ &$\cdots$ &$\cdots$  \\ 
       
       ACE (ESP, up)  &  $1.29 \pm 0.14$ &  $5.12 \pm 0.71$ & $0.46 \pm 0.07\: \mathrm{MeV}$ & $3.02 \pm 0.96$ & 
       $0.68 \pm 0.28$ & $8.74 \pm 3.40$ &$0.86 \pm 0.52$ & $1.33 \pm 0.41$  \\ 
       
       Solar Orbiter (ESP, down)  & $3.62 \pm 0.08$ &  $\cdots$ & $\cdots$ & $\cdots$ &  
       $3.81 \pm 0.06$& $\cdots$ &$\cdots$ &$\cdots$ \\ 
       
       ACE (ESP, down)  &  $2.17 \pm 0.10$ &  $13.48 \pm 13.71$ & $0.80 \pm 0.44$ & $4.80 \pm 3.14$&  
       $1.80 \pm 0.11$ &  $4.16 \pm 0.39$ &$0.35 \pm 0.05\: \mathrm{MeV}$&  $3.70 \pm 1.63$ \\ 
       \\
       & \multicolumn{4}{c}{$ \mathrm{Fe}$}&  \\ \cmidrule(lr){2-5} \\

       Solar Orbiter (SEP)  & $2.66 \pm 0.07 $  &  $\cdots$  & $\cdots$ & $\cdots$& \\
       
       ACE (SEP)  &$8.35 \pm 0.26$ &  $4.42 \pm 3.41$ & $0.25 \pm 0.22$ & $0.69 \pm 1.07$  \\
       
       Solar Orbiter (ESP)  & $3.17 \pm 0.06 $  &  $\cdots$  & $\cdots$ & $\cdots$\\
       
       ACE (ESP)  & $0.20 \pm 2.24$ &  $7.69 \pm 10.65$ & $0.39 \pm 1.08$ & $0.83 \pm 1.18$ \\
       
       Solar Orbiter (ESP, up)  & $3.01 \pm 0.08 $  &  $\cdots$  & $\cdots$ & $\cdots$  \\
       
       ACE (ESP, up)  & $8.13 \cdot 10^{-9} \pm 5.03$ &  $6.01 \pm 3.01$ & $0.22 \pm 0.13$ & $1.07 \pm 0.85$  \\
       
       Solar Orbiter (ESP, down)  & $3.63 \pm 0.06 $  &  $\cdots$  & $\cdots$ & $\cdots$  \\
       
       ACE (ESP, down)  & $1.28 \pm 1.75$ &  $11.41 \pm 55.81$ & $0.99 \pm 9.50$ & $0.96 \pm 2.10$ \\
       \\
    \end{tabular}
    \label{tabFit}
\end{sidewaystable*}

In this paper, we studied the radial evolution of ion species dependent acceleration using data from the 2022 March 10 ICME measured by both Solar Orbiter at 0.45 au and ACE at 1 au, with both spacecraft being radially aligned at the time of ICME transit. We compiled ion spectra for both spacecraft, separating the time intervals for the passage of the gradual SEP and ESP particles, to observe any changes in spectral shape. The observed spectra exhibited consistent differences for each time interval, with a lack of spectral breaks in Solar Orbiter spectra, and the higher energy portion of the ACE spectra having comparable average flux to the Solar Orbiter spectra (See Fig. \ref{FigResults}).

We came to a few key conclusions regarding the causes of these differences. For the spectra measured during the gradual SEP phase, the observed discrepancies are likely caused by the magnetic connectivity of the spacecraft, as certain portions of the shock can accelerate particles more efficiently, as well as an ion dropout. For the spectra measured during the ESP phase, we turned to the analysis of rigidity effects, focusing on the $\mathrm{Q/A}$ ordering of the spectral breaks seen at ACE. The sole misalignment of the $\mathrm{^{4}He}$ spectra relative to the other heavy ion spectra in rigidity space suggested an uncharacteristic overabundance of singly ionized, pickup helium at 1 au. We computed several rough reconstructions of the observed $\mathrm{^{4}He}$ spectra using an approximated composite spectra consisting of alpha particles ($\mathrm{He^{2+}}$) and pickup helium ($\mathrm{He^{+}}$), insinuating that the contribution of the latter was necessary to better align the $\mathrm{^{4}He}$ spectral shape with those of the other heavy ions. We speculate that the origin of this pool of $\mathrm{He^{+}}$ was a coincidentally passing SIR, as it would have been able to stream in pickup ions from the outer heliosphere where their concentration relative to $\mathrm{He^{2+}}$ is higher. The lack of spectral breaks in the spectra seen at Solar Orbiter is attributed to the influence of shock geometry, as surveying either the quasi-parallel or quasi-perpendicular portions of a shock has been shown to result in variations of observed spectral shape for the same event. Thus, we conclude that since both spacecraft did in fact view different portions of the shock as indicated by their measured shock thetas, the breaks present at 0.45 au were likely out of the observable energy range of Solar Orbiter rather than entirely absent.

To further validate the claims made in this paper, additional analysis will have to be conducted using multipoint observations of various ICMEs, preferably those that also interact with SIR events during or just prior to their transit.

\begin{acknowledgements}

Solar Orbiter is a mission of international cooperation between ESA and NASA, operated by ESA. Work on this study was funded by NASA SIS contract NNN06AA01C. MHW also acknowledges support from the Department of Defense (DoD) and the National Defense Science \& Engineering Graduate (NDSEG) Fellowship Program. RCA acknowledges support from NASA grants 80NSSC21K0733 and 80NSSC22K0993. A.K. acknowledges financial support from NASA HTM grant 80NSSC24K0071.

Observations from Solar Orbiter and ACE can be found at the Solar Orbiter Archive (SOAR; https://soar.esac.esa.int/soar/) and the ACE Science Center (https://izw1.caltech.edu/ACE/ASC/), respectively. We acknowledge the Community Coordinated Modeling Center (CCMC) at Goddard Space Flight Center for the use of the DONKI tool, https://kauai.ccmc.gsfc.nasa.gov/DONKI/. The EPD/Suprathermal Ion Spectrograph (SIS) is a European facility instrument funded by ESA under contract number SOL.ASTR.CON.00004.

This research has made use of Pythea v0.7.3, an open source and free Python package to reconstruct the 3D structure of CMEs and shock waves (Zenodo: https://doi.org/10.5281/zenodo.5713659).

\end{acknowledgements}

%
%

\bibliographystyle{aa}
\bibliography{main}

\begin{thebibliography}{94}
\expandafter\ifx\csname natexlab\endcsname\relax\def\natexlab#1{#1}\fi

\bibitem[{{Abraham-Shrauner} \& {Yun}(1976)}]{abyun1976}
{Abraham-Shrauner}, B. \& {Yun}, S.~H. 1976, \jgr, 81, 2097

\bibitem[{{Allen} {et~al.}(2021){Allen}, {Mason}, {Ho},
  {Rodr{\'\i}guez-Pacheco}, {Wimmer-Schweingruber}, {Andrews}, {Berger},
  {Boden}, {Cernuda}, {Espinosa Lara}, {Freiherr von Forstner},
  {G{\'o}mez-Herrero}, {Hayes}, {Kulkarni}, {Lees}, {Martin}, {Pacheco},
  {Polo}, {Prieto}, {Ravanbakhsh}, {S{\'a}nchez-Prieto}, {Schlemm}, {Seifert},
  {Terasa}, {Tyagi}, {Xu}, \& {Yedla}}]{allen2021}
{Allen}, R.~C., {Mason}, G.~M., {Ho}, G.~C., {et~al.} 2021, \aap, 656, L2

\bibitem[{{Band} {et~al.}(1993){Band}, {Matteson}, {Ford}, {Schaefer},
  {Palmer}, {Teegarden}, {Cline}, {Briggs}, {Paciesas}, {Pendleton}, {Fishman},
  {Kouveliotou}, {Meegan}, {Wilson}, \& {Lestrade}}]{band1993}
{Band}, D., {Matteson}, J., {Ford}, L., {et~al.} 1993, \apj, 413, 281

\bibitem[{{Brueckner} {et~al.}(1995){Brueckner}, {Howard}, {Koomen},
  {Korendyke}, {Michels}, {Moses}, {Socker}, {Dere}, {Lamy}, {Llebaria},
  {Bout}, {Schwenn}, {Simnett}, {Bedford}, \& {Eyles}}]{bru1995}
{Brueckner}, G.~E., {Howard}, R.~A., {Koomen}, M.~J., {et~al.} 1995, \solphys,
  162, 357

\bibitem[{{Burlaga} {et~al.}(2003){Burlaga}, {Berdichevsky}, {Gopalswamy},
  {Lepping}, \& {Zurbuchen}}]{burlaga2003}
{Burlaga}, L., {Berdichevsky}, D., {Gopalswamy}, N., {Lepping}, R., \&
  {Zurbuchen}, T. 2003, Journal of Geophysical Research (Space Physics), 108,
  1425

\bibitem[{{Chen} {et~al.}(2022){Chen}, {Giacalone}, \& {Guo}}]{chen2022}
{Chen}, X., {Giacalone}, J., \& {Guo}, F. 2022, \apj, 941, 23

\bibitem[{{Chotoo} {et~al.}(2000){Chotoo}, {Schwadron}, {Mason}, {Zurbuchen},
  {Gloeckler}, {Posner}, {Fisk}, {Galvin}, {Hamilton}, \& {Collier}}]{cho2000}
{Chotoo}, K., {Schwadron}, N.~A., {Mason}, G.~M., {et~al.} 2000, \jgr, 105,
  23107

\bibitem[{{Cohen} {et~al.}(2021){Cohen}, {Li}, {Mason}, {Shih}, \&
  {Wang}}]{cohen2021}
{Cohen}, C. M.~S., {Li}, G., {Mason}, G.~M., {Shih}, A.~Y., \& {Wang}, L. 2021,
  in Solar Physics and Solar Wind, ed. N.~E. {Raouafi} \& A.~{Vourlidas},
  Vol.~1, 133

\bibitem[{{Cohen} {et~al.}(2017){Cohen}, {Mason}, \& {Mewaldt}}]{cohen2017}
{Cohen}, C.~M.~S., {Mason}, G.~M., \& {Mewaldt}, R.~A. 2017, \apj, 843, 132

\bibitem[{{Cohen} {et~al.}(2003){Cohen}, {Mewaldt}, {Cummings}, {Leske},
  {Stone}, {von Rosenvinge}, \& {Wiedenbeck}}]{cohen2003}
{Cohen}, C.~M.~S., {Mewaldt}, R.~A., {Cummings}, A.~C., {et~al.} 2003, Advances
  in Space Research, 32, 2649

\bibitem[{{Cohen} {et~al.}(2005){Cohen}, {Stone}, {Mewaldt}, {Leske},
  {Cummings}, {Mason}, {Desai}, {von Rosenvinge}, \& {Wiedenbeck}}]{cohen2005}
{Cohen}, C.~M.~S., {Stone}, E.~C., {Mewaldt}, R.~A., {et~al.} 2005, Journal of
  Geophysical Research (Space Physics), 110, A09S16

\bibitem[{{Colburn} \& {Sonett}(1966)}]{colson1966}
{Colburn}, D.~S. \& {Sonett}, C.~P. 1966, \ssr, 5, 439

\bibitem[{{Desai} \& {Giacalone}(2016)}]{desai2016}
{Desai}, M. \& {Giacalone}, J. 2016, Living Reviews in Solar Physics, 13, 3

\bibitem[{{Desai} {et~al.}(2016){Desai}, {Mason}, {Dayeh}, {Ebert}, {McComas},
  {Li}, {Cohen}, {Mewaldt}, {Schwadron}, \& {Smith}}]{desai2016b}
{Desai}, M.~I., {Mason}, G.~M., {Dayeh}, M.~A., {et~al.} 2016, \apj, 828, 106

\bibitem[{{Desai} {et~al.}(2006){Desai}, {Mason}, {Gold}, {Krimigis}, {Cohen},
  {Mewaldt}, {Mazur}, \& {Dwyer}}]{desai2006}
{Desai}, M.~I., {Mason}, G.~M., {Gold}, R.~E., {et~al.} 2006, \apj, 649, 470

\bibitem[{{Desai} {et~al.}(2004){Desai}, {Mason}, {Wiedenbeck}, {Cohen},
  {Mazur}, {Dwyer}, {Gold}, {Krimigis}, {Hu}, {Smith}, \& {Skoug}}]{desai2004}
{Desai}, M.~I., {Mason}, G.~M., {Wiedenbeck}, M.~E., {et~al.} 2004, \apj, 611,
  1156

\bibitem[{{Domingo} {et~al.}(1995){Domingo}, {Fleck}, \& {Poland}}]{dom1995}
{Domingo}, V., {Fleck}, B., \& {Poland}, A.~I. 1995, \solphys, 162, 1

\bibitem[{{Dresing} {et~al.}(2023){Dresing}, {Rodr{\'\i}guez-Garc{\'\i}a},
  {Jebaraj}, {Warmuth}, {Wallace}, {Balmaceda}, {Podladchikova}, {Strauss},
  {Kouloumvakos}, {Palmroos}, {Krupar}, {Gieseler}, {Xu}, {Mitchell}, {Cohen},
  {de Nolfo}, {Palmerio}, {Carcaboso}, {Kilpua}, {Trotta}, {Auster},
  {Asvestari}, {da Silva}, {Dr{\"o}ge}, {Getachew}, {G{\'o}mez-Herrero},
  {Grande}, {Heyner}, {Holmstr{\"o}m}, {Huovelin}, {Kartavykh}, {Laurenza},
  {Lee}, {Mason}, {Maksimovic}, {Mieth}, {Murakami}, {Oleynik}, {Pinto},
  {Pulupa}, {Richter}, {Rodr{\'\i}guez-Pacheco}, {S{\'a}nchez-Cano},
  {Schuller}, {Ueno}, {Vainio}, {Vecchio}, {Veronig}, \&
  {Wijsen}}]{dresing2023}
{Dresing}, N., {Rodr{\'\i}guez-Garc{\'\i}a}, L., {Jebaraj}, I.~C., {et~al.}
  2023, \aap, 674, A105

\bibitem[{{Drury}(1983)}]{dru1983}
{Drury}, L.~O. 1983, Reports on Progress in Physics, 46, 973

\bibitem[{{Ellison} \& {Ramaty}(1985)}]{er1985}
{Ellison}, D.~C. \& {Ramaty}, R. 1985, \apj, 298, 400

\bibitem[{{Garc{\'\i}a-Rigo} {et~al.}(2016){Garc{\'\i}a-Rigo}, {N{\'u}{\~n}ez},
  {Qahwaji}, {Ashamari}, {Jiggens}, {P{\'e}rez}, {Hern{\'a}ndez-Pajares}, \&
  {Hilgers}}]{garcia2016}
{Garc{\'\i}a-Rigo}, A., {N{\'u}{\~n}ez}, M., {Qahwaji}, R., {et~al.} 2016,
  Journal of Space Weather and Space Climate, 6, A28

\bibitem[{{Giacalone} {et~al.}(2021){Giacalone}, {Burgess}, {Bale}, {Desai},
  {Mitchell}, {Lario}, {Chen}, {Christian}, {de Nolfo}, {Hill}, {Matthaeus},
  {McComas}, {McNutt}, {Mitchell}, {Roelof}, {Schwadron}, {Getachew}, \&
  {Joyce}}]{giac2021}
{Giacalone}, J., {Burgess}, D., {Bale}, S.~D., {et~al.} 2021, \apj, 921, 102

\bibitem[{{Giacalone} {et~al.}(2020){Giacalone}, {Mitchell}, {Allen}, {Hill},
  {McNutt}, {Szalay}, {Desai}, {Rouillard}, {Kouloumvakos}, {McComas},
  {Christian}, {Schwadron}, {Wiedenbeck}, {Bale}, {Brown}, {Case}, {Chen},
  {Cohen}, {Joyce}, {Kasper}, {Klein}, {Korreck}, {Larson}, {Livi}, {Leske},
  {MacDowall}, {Matthaeus}, {Mewaldt}, {Nieves-Chinchilla}, {Pulupa}, {Roelof},
  {Stevens}, {Szabo}, \& {Whittlesey}}]{giac2020}
{Giacalone}, J., {Mitchell}, D.~G., {Allen}, R.~C., {et~al.} 2020, \apjs, 246,
  29

\bibitem[{{Gieseler} {et~al.}(2023){Gieseler}, {Dresing}, {Palmroos}, {Freiherr
  von Forstner}, {Price}, {Vainio}, {Kouloumvakos},
  {Rodr{\'\i}guez-Garc{\'\i}a}, {Trotta}, {G{\'e}not}, {Masson}, {Roth}, \&
  {Veronig}}]{gies2023}
{Gieseler}, J., {Dresing}, N., {Palmroos}, C., {et~al.} 2023, Frontiers in
  Astronomy and Space Sciences, 9, 384

\bibitem[{{Gold} {et~al.}(1998){Gold}, {Krimigis}, {Hawkins}, {Haggerty},
  {Lohr}, {Fiore}, {Armstrong}, {Holland}, \& {Lanzerotti}}]{gold1998}
{Gold}, R.~E., {Krimigis}, S.~M., {Hawkins}, S.~E., I., {et~al.} 1998, \ssr,
  86, 541

\bibitem[{{Gopalswamy} {et~al.}(2018){Gopalswamy}, {M{\"a}kel{\"a}}, {Akiyama},
  {Yashiro}, {Xie}, \& {Thakur}}]{gopal2018}
{Gopalswamy}, N., {M{\"a}kel{\"a}}, P., {Akiyama}, S., {et~al.} 2018, Journal
  of Atmospheric and Solar-Terrestrial Physics, 179, 225

\bibitem[{{Gopalswamy} {et~al.}(2014){Gopalswamy}, {Xie}, {Akiyama},
  {M{\"a}kel{\"a}}, \& {Yashiro}}]{gopal2014}
{Gopalswamy}, N., {Xie}, H., {Akiyama}, S., {M{\"a}kel{\"a}}, P.~A., \&
  {Yashiro}, S. 2014, Earth, Planets and Space, 66, 104

\bibitem[{{Gopalswamy} {et~al.}(2016){Gopalswamy}, {Yashiro}, {Thakur},
  {M{\"a}kel{\"a}}, {Xie}, \& {Akiyama}}]{gopal2016}
{Gopalswamy}, N., {Yashiro}, S., {Thakur}, N., {et~al.} 2016, \apj, 833, 216

\bibitem[{{Gosling} {et~al.}(1978){Gosling}, {Asbridge}, {Bame}, \&
  {Feldman}}]{gosling1978}
{Gosling}, J.~T., {Asbridge}, J.~R., {Bame}, S.~J., \& {Feldman}, W.~C. 1978,
  \jgr, 83, 1401

\bibitem[{{Gruesbeck} {et~al.}(2011){Gruesbeck}, {Lepri}, {Zurbuchen}, \&
  {Antiochos}}]{gru2011}
{Gruesbeck}, J.~R., {Lepri}, S.~T., {Zurbuchen}, T.~H., \& {Antiochos}, S.~K.
  2011, \apj, 730, 103

\bibitem[{{Ho} {et~al.}(2022){Ho}, {Mason}, {Allen}, {Wimmer-Schweingruber},
  {Rodr{\'\i}guez-Pacheco}, \& {G{\'o}mez-Herrero}}]{ho2022}
{Ho}, G.~C., {Mason}, G.~M., {Allen}, R.~C., {et~al.} 2022, Frontiers in
  Astronomy and Space Sciences, 9, 939799

\bibitem[{{Holzer}(1972)}]{holzer1972}
{Holzer}, T.~E. 1972, \jgr, 77, 5407

\bibitem[{{Horbury} {et~al.}(2020){Horbury}, {O'Brien}, {Carrasco Blazquez},
  {Bendyk}, {Brown}, {Hudson}, {Evans}, {Oddy}, {Carr}, {Beek}, {Cupido},
  {Bhattacharya}, {Dominguez}, {Matthews}, {Myklebust}, {Whiteside}, {Bale},
  {Baumjohann}, {Burgess}, {Carbone}, {Cargill}, {Eastwood}, {Erd{\"o}s},
  {Fletcher}, {Forsyth}, {Giacalone}, {Glassmeier}, {Goldstein}, {Hoeksema},
  {Lockwood}, {Magnes}, {Maksimovic}, {Marsch}, {Matthaeus}, {Murphy},
  {Nakariakov}, {Owen}, {Owens}, {Rodriguez-Pacheco}, {Richter}, {Riley},
  {Russell}, {Schwartz}, {Vainio}, {Velli}, {Vennerstrom}, {Walsh},
  {Wimmer-Schweingruber}, {Zank}, {M{\"u}ller}, {Zouganelis}, \&
  {Walsh}}]{horb2020}
{Horbury}, T.~S., {O'Brien}, H., {Carrasco Blazquez}, I., {et~al.} 2020, \aap,
  642, A9

\bibitem[{{Howard} {et~al.}(2008){Howard}, {Moses}, {Vourlidas}, {Newmark},
  {Socker}, {Plunkett}, {Korendyke}, {Cook}, {Hurley}, {Davila}, {Thompson},
  {St Cyr}, {Mentzell}, {Mehalick}, {Lemen}, {Wuelser}, {Duncan}, {Tarbell},
  {Wolfson}, {Moore}, {Harrison}, {Waltham}, {Lang}, {Davis}, {Eyles},
  {Mapson-Menard}, {Simnett}, {Halain}, {Defise}, {Mazy}, {Rochus}, {Mercier},
  {Ravet}, {Delmotte}, {Auchere}, {Delaboudiniere}, {Bothmer}, {Deutsch},
  {Wang}, {Rich}, {Cooper}, {Stephens}, {Maahs}, {Baugh}, {McMullin}, \&
  {Carter}}]{howard2008}
{Howard}, R.~A., {Moses}, J.~D., {Vourlidas}, A., {et~al.} 2008, \ssr, 136, 67

\bibitem[{{Isenberg}(1986)}]{isen1986}
{Isenberg}, P.~A. 1986, \jgr, 91, 9965

\bibitem[{{Jian} {et~al.}(2006){Jian}, {Russell}, {Luhmann}, \&
  {Skoug}}]{jian2006}
{Jian}, L., {Russell}, C.~T., {Luhmann}, J.~G., \& {Skoug}, R.~M. 2006,
  \solphys, 239, 337

\bibitem[{{Jian} {et~al.}(2008){Jian}, {Russell}, {Luhmann}, {Skoug}, \&
  {Steinberg}}]{jian2008}
{Jian}, L.~K., {Russell}, C.~T., {Luhmann}, J.~G., {Skoug}, R.~M., \&
  {Steinberg}, J.~T. 2008, \solphys, 250, 375

\bibitem[{{Khabibrakhmanov} {et~al.}(1996){Khabibrakhmanov}, {Summers}, {Zank},
  \& {Pauls}}]{khab1996}
{Khabibrakhmanov}, I.~K., {Summers}, D., {Zank}, G.~P., \& {Pauls}, H.~L. 1996,
  \apj, 469, 921

\bibitem[{{Kilpua} {et~al.}(2015){Kilpua}, {Lumme}, {Andreeova}, {Isavnin}, \&
  {Koskinen}}]{kil2015}
{Kilpua}, E.~K.~J., {Lumme}, E., {Andreeova}, K., {Isavnin}, A., \& {Koskinen},
  H.~E.~J. 2015, Journal of Geophysical Research (Space Physics), 120, 4112

\bibitem[{{Klecker} {et~al.}(2007{\natexlab{a}}){Klecker}, {M{\"o}bius}, \&
  {Popecki}}]{klec2007a}
{Klecker}, B., {M{\"o}bius}, E., \& {Popecki}, M.~A. 2007{\natexlab{a}}, \ssr,
  130, 273

\bibitem[{{Klecker} {et~al.}(2007{\natexlab{b}}){Klecker}, {M{\"o}bius}, \&
  {Popecki}}]{klec2007b}
{Klecker}, B., {M{\"o}bius}, E., \& {Popecki}, M.~A. 2007{\natexlab{b}}, in
  Solar Dynamics and Its Effects on the Heliosphere and Earth, ed. D.~N.
  {Baker}, B.~{Klecker}, S.~J. {Schwartz}, R.~{Schwenn}, \& R.~{von Steiger},
  Vol.~22, 289--301

\bibitem[{{Ko} {et~al.}(1999){Ko}, {Gloeckler}, {Cohen}, \& {Galvin}}]{ko1999}
{Ko}, Y.-K., {Gloeckler}, G., {Cohen}, C. M.~S., \& {Galvin}, A.~B. 1999, \jgr,
  104, 17005

\bibitem[{{Kollhoff} {et~al.}(2021){Kollhoff}, {Kouloumvakos}, {Lario},
  {Dresing}, {G{\'o}mez-Herrero}, {Rodr{\'\i}guez-Garc{\'\i}a}, {Malandraki},
  {Richardson}, {Posner}, {Klein}, {Pacheco}, {Klassen}, {Heber}, {Cohen},
  {Laitinen}, {Cernuda}, {Dalla}, {Espinosa Lara}, {Vainio}, {K{\"o}berle},
  {K{\"u}hl}, {Xu}, {Berger}, {Eldrum}, {Br{\"u}dern}, {Laurenza}, {Kilpua},
  {Aran}, {Rouillard}, {Bu{\v{c}}{\'\i}k}, {Wijsen}, {Pomoell},
  {Wimmer-Schweingruber}, {Martin}, {B{\"o}ttcher}, {Freiherr von Forstner},
  {Terasa}, {Boden}, {Kulkarni}, {Ravanbakhsh}, {Yedla}, {Janitzek},
  {Rodr{\'\i}guez-Pacheco}, {Prieto Mateo}, {S{\'a}nchez Prieto}, {Parra
  Espada}, {Rodr{\'\i}guez Polo}, {Mart{\'\i}nez Hell{\'\i}n}, {Carcaboso},
  {Mason}, {Ho}, {Allen}, {Bruce Andrews}, {Schlemm}, {Seifert}, {Tyagi},
  {Lees}, {Hayes}, {Bale}, {Krupar}, {Horbury}, {Angelini}, {Evans}, {O'Brien},
  {Maksimovic}, {Khotyaintsev}, {Vecchio}, {Steinvall}, \&
  {Asvestari}}]{koll2021}
{Kollhoff}, A., {Kouloumvakos}, A., {Lario}, D., {et~al.} 2021, \aap, 656, A20

\bibitem[{{Kouloumvakos} {et~al.}(2022){Kouloumvakos},
  {Rodr{\'\i}guez-Garc{\'\i}a}, {Gieseler}, {Price}, {Vourlidas}, \&
  {Vainio}}]{koul2022}
{Kouloumvakos}, A., {Rodr{\'\i}guez-Garc{\'\i}a}, L., {Gieseler}, J., {et~al.}
  2022, Frontiers in Astronomy and Space Sciences, 9, 974137

\bibitem[{{Kucharek} {et~al.}(2003){Kucharek}, {M{\"o}Bius}, {Li}, {Farrugia},
  {Popecki}, {Galvin}, {Klecker}, {Hilchenbach}, \& {Bochsler}}]{kuch2003}
{Kucharek}, H., {M{\"o}Bius}, E., {Li}, W., {et~al.} 2003, Journal of
  Geophysical Research (Space Physics), 108, 8040

\bibitem[{{Laming} {et~al.}(2023){Laming}, {Provornikova}, \& {Ko}}]{lam2023}
{Laming}, J.~M., {Provornikova}, E., \& {Ko}, Y.-K. 2023, \apj, 954, 145

\bibitem[{{Lario} {et~al.}(2003){Lario}, {Ho}, {Decker}, {Roelof}, {Desai}, \&
  {Smith}}]{lario2003}
{Lario}, D., {Ho}, G.~C., {Decker}, R.~B., {et~al.} 2003, in American Institute
  of Physics Conference Series, Vol. 679, Solar Wind Ten, ed. M.~{Velli},
  R.~{Bruno}, F.~{Malara}, \& B.~{Bucci}, 640--643

\bibitem[{{le Roux} {et~al.}(2015){le Roux}, {Zank}, {Webb}, \&
  {Khabarova}}]{leroux2015}
{le Roux}, J.~A., {Zank}, G.~P., {Webb}, G.~M., \& {Khabarova}, O. 2015, \apj,
  801, 112

\bibitem[{{Lee}(2000)}]{lee2000}
{Lee}, M.~A. 2000, in American Institute of Physics Conference Series, Vol.
  528, Acceleration and Transport of Energetic Particles Observed in the
  Heliosphere, ed. R.~A. {Mewaldt}, J.~R. {Jokipii}, M.~A. {Lee},
  E.~{M{\"o}bius}, \& T.~H. {Zurbuchen}, 3--18

\bibitem[{{Li} {et~al.}(2009){Li}, {Zank}, {Verkhoglyadova}, {Mewaldt},
  {Cohen}, {Mason}, \& {Desai}}]{li2009}
{Li}, G., {Zank}, G.~P., {Verkhoglyadova}, O., {et~al.} 2009, \apj, 702, 998

\bibitem[{{Liu} {et~al.}(2020){Liu}, {Wang}, {Wimmer-Schweingruber}, {Krucker},
  \& {Mason}}]{liu2020}
{Liu}, Z., {Wang}, L., {Wimmer-Schweingruber}, R.~F., {Krucker}, S., \&
  {Mason}, G.~M. 2020, Journal of Geophysical Research (Space Physics), 125,
  e28702

\bibitem[{{Luhmann} {et~al.}(2020){Luhmann}, {Gopalswamy}, {Jian}, \&
  {Lugaz}}]{luhm2020}
{Luhmann}, J.~G., {Gopalswamy}, N., {Jian}, L.~K., \& {Lugaz}, N. 2020,
  \solphys, 295, 61

\bibitem[{{Maclean} {et~al.}(2014){Maclean}, {Pagnozzi}, \& {Biggs}}]{mac2014}
{Maclean}, C., {Pagnozzi}, D., \& {Biggs}, J. 2014, IEEE Transactions on
  Aerospace Electronic Systems, 50, 2129

\bibitem[{{Mason} {et~al.}(2021){Mason}, {Cohen}, {Ho}, {Mitchell}, {Allen},
  {Hill}, {Andrews}, {Berger}, {Boden}, {B{\"o}ttcher}, {Cernuda}, {Christian},
  {Cummings}, {Davis}, {Desai}, {de Nolfo}, {Eldrum}, {Elftmann}, {Kollhoff},
  {Giacalone}, {G{\'o}mez-Herrero}, {Hayes}, {Janitzek}, {Joyce}, {Korth},
  {K{\"u}hl}, {Kulkarni}, {Labrador}, {Espinosa Lara}, {Lees}, {Leske}, {Mall},
  {Martin}, {Mart{\'\i}nez Hell{\'\i}n}, {Matthaeus}, {McComas}, {McNutt},
  {Mewaldt}, {Mitchell}, {Pacheco}, {Parra Espada}, {Prieto}, {Rankin},
  {Ravanbakhsh}, {Rodr{\'\i}guez-Pacheco}, {Rodr{\'\i}guez Polo}, {Roelof},
  {S{\'a}nchez-Prieto}, {Schlemm}, {Schwadron}, {Seifert}, {Stone}, {Szalay},
  {Terasa}, {Tyagi}, {Freiherr von Forstner}, {Wiedenbeck},
  {Wimmer-Schweingruber}, {Xu}, \& {Yedla}}]{mason2021}
{Mason}, G.~M., {Cohen}, C.~M.~S., {Ho}, G.~C., {et~al.} 2021, \aap, 656, L12

\bibitem[{{Mason} {et~al.}(1998){Mason}, {Gold}, {Krimigis}, {Mazur},
  {Andrews}, {Daley}, {Dwyer}, {Heuerman}, {James}, {Kennedy}, {Lefevere},
  {Malcolm}, {Tossman}, \& {Walpole}}]{mason1998}
{Mason}, G.~M., {Gold}, R.~E., {Krimigis}, S.~M., {et~al.} 1998, \ssr, 86, 409

\bibitem[{{Mason} {et~al.}(2008){Mason}, {Leske}, {Desai}, {Cohen}, {Dwyer},
  {Mazur}, {Mewaldt}, {Gold}, \& {Krimigis}}]{mason2008}
{Mason}, G.~M., {Leske}, R.~A., {Desai}, M.~I., {et~al.} 2008, \apj, 678, 1458

\bibitem[{{Mason} {et~al.}(2012){Mason}, {Li}, {Cohen}, {Desai}, {Haggerty},
  {Leske}, {Mewaldt}, \& {Zank}}]{mason2012}
{Mason}, G.~M., {Li}, G., {Cohen}, C.~M.~S., {et~al.} 2012, \apj, 761, 104

\bibitem[{{Mays} {et~al.}(2015){Mays}, {Taktakishvili}, {Pulkkinen},
  {MacNeice}, {Rast{\"a}tter}, {Odstrcil}, {Jian}, {Richardson}, {LaSota},
  {Zheng}, \& {Kuznetsova}}]{mays2015}
{Mays}, M.~L., {Taktakishvili}, A., {Pulkkinen}, A., {et~al.} 2015, \solphys,
  290, 1775

\bibitem[{{McComas} {et~al.}(1998){McComas}, {Bame}, {Barker}, {Feldman},
  {Phillips}, {Riley}, \& {Griffee}}]{mc1998}
{McComas}, D.~J., {Bame}, S.~J., {Barker}, P., {et~al.} 1998, \ssr, 86, 563

\bibitem[{{Mewaldt} {et~al.}(2005{\natexlab{a}}){Mewaldt}, {Cohen}, {Labrador},
  {Leske}, {Mason}, {Desai}, {Looper}, {Mazur}, {Selesnick}, \&
  {Haggerty}}]{mewaldt2005b}
{Mewaldt}, R.~A., {Cohen}, C.~M.~S., {Labrador}, A.~W., {et~al.}
  2005{\natexlab{a}}, Journal of Geophysical Research (Space Physics), 110,
  A09S18

\bibitem[{{Mewaldt} {et~al.}(2005{\natexlab{b}}){Mewaldt}, {Cohen}, {Mason},
  {Labrador}, {Looper}, {Haggerty}, {Maclennan}, {Cummings}, {Desai}, {Leske},
  {Li}, {Mazur}, {Stone}, \& {Wiedenbeck}}]{mewaldt2005a}
{Mewaldt}, R.~A., {Cohen}, C.~M.~S., {Mason}, G.~M., {et~al.}
  2005{\natexlab{b}}, in American Institute of Physics Conference Series, Vol.
  781, The Physics of Collisionless Shocks: 4th Annual IGPP International
  Astrophysics Conference, ed. G.~{Li}, G.~P. {Zank}, \& C.~T. {Russell},
  227--232

\bibitem[{{Mewaldt} {et~al.}(2001){Mewaldt}, {Mason}, {Gloeckler}, {Christian},
  {Cohen}, {Cummings}, {Davis}, {Dwyer}, {Gold}, {Krimigis}, {Leske}, {Mazur},
  {Stone}, {von Rosenvinge}, {Wiedenbeck}, \& {Zurbuchen}}]{mew2001}
{Mewaldt}, R.~A., {Mason}, G.~M., {Gloeckler}, G., {et~al.} 2001, in American
  Institute of Physics Conference Series, Vol. 598, Joint SOHO/ACE workshop
  ``Solar and Galactic Composition'', ed. R.~F. {Wimmer-Schweingruber},
  165--170

\bibitem[{{M{\"o}bius} {et~al.}(2000){M{\"o}bius}, {Klecker}, {Popecki},
  {Morris}, {Mason}, {Stone}, {Bogdanov}, {Dwyer}, {Galvin}, {Heirtzler},
  {Hovestadt}, {Kistler}, \& {Siren}}]{mobius2000}
{M{\"o}bius}, E., {Klecker}, B., {Popecki}, M.~A., {et~al.} 2000, in American
  Institute of Physics Conference Series, Vol. 528, Acceleration and Transport
  of Energetic Particles Observed in the Heliosphere, ed. R.~A. {Mewaldt},
  J.~R. {Jokipii}, M.~A. {Lee}, E.~{M{\"o}bius}, \& T.~H. {Zurbuchen} (AIP),
  131--134

\bibitem[{{Morris} {et~al.}(2001){Morris}, {M{\"o}bius}, {Lee}, {Popecki},
  {Klecker}, {Kistler}, \& {Galvin}}]{morris2001}
{Morris}, D., {M{\"o}bius}, E., {Lee}, M.~A., {et~al.} 2001, in American
  Institute of Physics Conference Series, Vol. 598, Joint SOHO/ACE workshop
  ``Solar and Galactic Composition'', ed. R.~F. {Wimmer-Schweingruber},
  201--204

\bibitem[{{M{\"u}ller} {et~al.}(2020){M{\"u}ller}, {St. Cyr}, {Zouganelis},
  {Gilbert}, {Marsden}, {Nieves-Chinchilla}, {Antonucci}, {Auch{\`e}re},
  {Berghmans}, {Horbury}, {Howard}, {Krucker}, {Maksimovic}, {Owen}, {Rochus},
  {Rodriguez-Pacheco}, {Romoli}, {Solanki}, {Bruno}, {Carlsson}, {Fludra},
  {Harra}, {Hassler}, {Livi}, {Louarn}, {Peter}, {Sch{\"u}hle}, {Teriaca}, {del
  Toro Iniesta}, {Wimmer-Schweingruber}, {Marsch}, {Velli}, {De Groof},
  {Walsh}, \& {Williams}}]{muller2020}
{M{\"u}ller}, D., {St. Cyr}, O.~C., {Zouganelis}, I., {et~al.} 2020, \aap, 642,
  A1

\bibitem[{{Odstrcil} {et~al.}(2020){Odstrcil}, {Mays}, {Hess}, {Jones},
  {Henney}, \& {Arge}}]{od2020}
{Odstrcil}, D., {Mays}, M.~L., {Hess}, P., {et~al.} 2020, \apjs, 246, 73

\bibitem[{{Odstrcil} {et~al.}(2004){Odstrcil}, {Riley}, \& {Zhao}}]{od2004}
{Odstrcil}, D., {Riley}, P., \& {Zhao}, X.~P. 2004, Journal of Geophysical
  Research (Space Physics), 109, A02116

\bibitem[{{Owen} {et~al.}(2020){Owen}, {Bruno}, {Livi}, {Louarn}, {Al Janabi},
  {Allegrini}, {Amoros}, {Baruah}, {Barthe}, {Berthomier}, {Bordon},
  {Brockley-Blatt}, {Brysbaert}, {Capuano}, {Collier}, {DeMarco}, {Fedorov},
  {Ford}, {Fortunato}, {Fratter}, {Galvin}, {Hancock}, {Heirtzler}, {Kataria},
  {Kistler}, {Lepri}, {Lewis}, {Loeffler}, {Marty}, {Mathon}, {Mayall}, {Mele},
  {Ogasawara}, {Orlandi}, {Pacros}, {Penou}, {Persyn}, {Petiot}, {Phillips},
  {P{\v{r}}ech}, {Raines}, {Reden}, {Rouillard}, {Rousseau}, {Rubiella},
  {Seran}, {Spencer}, {Thomas}, {Trevino}, {Verscharen}, {Wurz}, {Alapide},
  {Amoruso}, {Andr{\'e}}, {Anekallu}, {Arciuli}, {Arnett}, {Ascolese},
  {Bancroft}, {Bland}, {Brysch}, {Calvanese}, {Castronuovo},
  {{\v{C}}erm{\'a}k}, {Chornay}, {Clemens}, {Coker}, {Collinson}, {D'Amicis},
  {Dandouras}, {Darnley}, {Davies}, {Davison}, {De Los Santos}, {Devoto},
  {Dirks}, {Edlund}, {Fazakerley}, {Ferris}, {Frost}, {Fruit}, {Garat},
  {G{\'e}not}, {Gibson}, {Gilbert}, {de Giosa}, {Gradone}, {Hailey}, {Horbury},
  {Hunt}, {Jacquey}, {Johnson}, {Lavraud}, {Lawrenson}, {Leblanc}, {Lockhart},
  {Maksimovic}, {Malpus}, {Marcucci}, {Mazelle}, {Monti}, {Myers}, {Nguyen},
  {Rodriguez-Pacheco}, {Phillips}, {Popecki}, {Rees}, {Rogacki}, {Ruane},
  {Rust}, {Salatti}, {Sauvaud}, {Stakhiv}, {Stange}, {Stubbs}, {Taylor},
  {Techer}, {Terrier}, {Thibodeaux}, {Urdiales}, {Varsani}, {Walsh}, {Watson},
  {Wheeler}, {Willis}, {Wimmer-Schweingruber}, {Winter}, {Yardley}, \&
  {Zouganelis}}]{owen2020}
{Owen}, C.~J., {Bruno}, R., {Livi}, S., {et~al.} 2020, \aap, 642, A16

\bibitem[{{Palmerio} {et~al.}(2024){Palmerio}, {Carcaboso}, {Khoo}, {Salman},
  {S{\'a}nchez-Cano}, {Lynch}, {Rivera}, {Pal}, {Nieves-Chinchilla}, {Weiss},
  {Lario}, {Mieth}, {Heyner}, {Stevens}, {Romeo}, {Zhukov}, {Rodriguez}, {Lee},
  {Cohen}, {Rodr{\'\i}guez-Garc{\'\i}a}, {Whittlesey}, {Dresing}, {Oleynik},
  {Jebaraj}, {Fischer}, {Schmid}, {Richter}, {Auster}, {Fraschetti}, \&
  {Mierla}}]{palm2024}
{Palmerio}, E., {Carcaboso}, F., {Khoo}, L.~Y., {et~al.} 2024, \apj, 963, 108

\bibitem[{{Palmerio} {et~al.}(2022){Palmerio}, {Lee}, {Mays}, {Luhmann},
  {Lario}, {S{\'a}nchez-Cano}, {Richardson}, {Vainio}, {Stevens}, {Cohen},
  {Steinvall}, {M{\"o}stl}, {Weiss}, {Nieves-Chinchilla}, {Li}, {Larson},
  {Heyner}, {Bale}, {Galvin}, {Holmstr{\"o}m}, {Khotyaintsev}, {Maksimovic}, \&
  {Mitrofanov}}]{palm2022}
{Palmerio}, E., {Lee}, C.~O., {Mays}, M.~L., {et~al.} 2022, Space Weather, 20,
  e2021SW002993

\bibitem[{{Pizzo}(1978)}]{pizzo1978}
{Pizzo}, V. 1978, \jgr, 83, 5563

\bibitem[{{Rakowski} {et~al.}(2007){Rakowski}, {Laming}, \&
  {Lepri}}]{rakowski2007}
{Rakowski}, C.~E., {Laming}, J.~M., \& {Lepri}, S.~T. 2007, \apj, 667, 602

\bibitem[{{Reames}(2020)}]{reames2020}
{Reames}, D.~V. 2020, \solphys, 295, 113

\bibitem[{{Richardson}(2004)}]{rich2004}
{Richardson}, I.~G. 2004, \ssr, 111, 267

\bibitem[{{Riordan} \& {Pe'er}(2019)}]{riordan2019}
{Riordan}, J.~D. \& {Pe'er}, A. 2019, \apj, 873, 13

\bibitem[{{Rodr{\'\i}guez-Garc{\'\i}a}
  {et~al.}(2023){Rodr{\'\i}guez-Garc{\'\i}a}, {Balmaceda}, {G{\'o}mez-Herrero},
  {Kouloumvakos}, {Dresing}, {Lario}, {Zouganelis}, {Fedeli}, {Espinosa Lara},
  {Cernuda}, {Ho}, {Wimmer-Schweingruber}, \&
  {Rodr{\'\i}guez-Pacheco}}]{rod2023}
{Rodr{\'\i}guez-Garc{\'\i}a}, L., {Balmaceda}, L.~A., {G{\'o}mez-Herrero}, R.,
  {et~al.} 2023, \aap, 674, A145

\bibitem[{{Rodr{\'\i}guez-Pacheco} {et~al.}(2020){Rodr{\'\i}guez-Pacheco},
  {Wimmer-Schweingruber}, {Mason}, {Ho}, {S{\'a}nchez-Prieto}, {Prieto},
  {Mart{\'\i}n}, {Seifert}, {Andrews}, {Kulkarni}, {Panitzsch}, {Boden},
  {B{\"o}ttcher}, {Cernuda}, {Elftmann}, {Espinosa Lara}, {G{\'o}mez-Herrero},
  {Terasa}, {Almena}, {Begley}, {B{\"o}hm}, {Blanco}, {Boogaerts}, {Carrasco},
  {Castillo}, {da Silva Fari{\~n}a}, {de Manuel Gonz{\'a}lez}, {Drews},
  {Dupont}, {Eldrum}, {Gordillo}, {Guti{\'e}rrez}, {Haggerty}, {Hayes},
  {Heber}, {Hill}, {J{\"u}ngling}, {Kerem}, {Knierim}, {K{\"o}hler}, {Kolbe},
  {Kulemzin}, {Lario}, {Lees}, {Liang}, {Mart{\'\i}nez Hell{\'\i}n}, {Meziat},
  {Montalvo}, {Nelson}, {Parra}, {Paspirgilis}, {Ravanbakhsh}, {Richards},
  {Rodr{\'\i}guez-Polo}, {Russu}, {S{\'a}nchez}, {Schlemm}, {Schuster},
  {Seimetz}, {Steinhagen}, {Tammen}, {Tyagi}, {Varela}, {Yedla}, {Yu},
  {Agueda}, {Aran}, {Horbury}, {Klecker}, {Klein}, {Kontar}, {Krucker},
  {Maksimovic}, {Malandraki}, {Owen}, {Pacheco}, {Sanahuja}, {Vainio},
  {Connell}, {Dalla}, {Dr{\"o}ge}, {Gevin}, {Gopalswamy}, {Kartavykh},
  {Kudela}, {Limousin}, {Makela}, {Mann}, {{\"O}nel}, {Posner}, {Ryan},
  {Soucek}, {Hofmeister}, {Vilmer}, {Walsh}, {Wang}, {Wiedenbeck}, {Wirth}, \&
  {Zong}}]{rodr2020}
{Rodr{\'\i}guez-Pacheco}, J., {Wimmer-Schweingruber}, R.~F., {Mason}, G.~M.,
  {et~al.} 2020, \aap, 642, A7

\bibitem[{{Schwadron} {et~al.}(2015){Schwadron}, {Lee}, {Gorby}, {Lugaz},
  {Spence}, {Desai}, {T{\"o}r{\"o}k}, {Downs}, {Linker}, {Lionello},
  {Miki{\'c}}, {Riley}, {Giacalone}, {Jokipii}, {Kota}, \& {Kozarev}}]{sch2015}
{Schwadron}, N.~A., {Lee}, M.~A., {Gorby}, M., {et~al.} 2015, \apj, 810, 97

\bibitem[{{Shen} {et~al.}(2017){Shen}, {Chi}, {Wang}, {Xu}, \&
  {Wang}}]{shen2017}
{Shen}, C., {Chi}, Y., {Wang}, Y., {Xu}, M., \& {Wang}, S. 2017, Journal of
  Geophysical Research (Space Physics), 122, 5931

\bibitem[{{Shen} {et~al.}(2018){Shen}, {Xu}, {Wang}, {Chi}, \&
  {Luo}}]{shen2018}
{Shen}, C., {Xu}, M., {Wang}, Y., {Chi}, Y., \& {Luo}, B. 2018, \apj, 861, 28

\bibitem[{{Smith} {et~al.}(1998){Smith}, {L'Heureux}, {Ness}, {Acu{\~n}a},
  {Burlaga}, \& {Scheifele}}]{smith1998}
{Smith}, C.~W., {L'Heureux}, J., {Ness}, N.~F., {et~al.} 1998, \ssr, 86, 613

\bibitem[{{Smith} \& {Wolfe}(1976)}]{smith1976}
{Smith}, E.~J. \& {Wolfe}, J.~H. 1976, \grl, 3, 137

\bibitem[{{Stone} {et~al.}(1998){Stone}, {Frandsen}, {Mewaldt}, {Christian},
  {Margolies}, {Ormes}, \& {Snow}}]{stone1998}
{Stone}, E.~C., {Frandsen}, A.~M., {Mewaldt}, R.~A., {et~al.} 1998, \ssr, 86, 1

\bibitem[{{Tsurutani} {et~al.}(2003){Tsurutani}, {Wu}, {Zhang}, \&
  {Dryer}}]{tsu2003}
{Tsurutani}, B., {Wu}, S.~T., {Zhang}, T.~X., \& {Dryer}, M. 2003, \aap, 412,
  293

\bibitem[{{Tylka} {et~al.}(2000){Tylka}, {Boberg}, {McGuire}, {Ng}, \&
  {Reames}}]{ty2000}
{Tylka}, A.~J., {Boberg}, P.~R., {McGuire}, R.~E., {Ng}, C.~K., \& {Reames},
  D.~V. 2000, in American Institute of Physics Conference Series, Vol. 528,
  Acceleration and Transport of Energetic Particles Observed in the
  Heliosphere, ed. R.~A. {Mewaldt}, J.~R. {Jokipii}, M.~A. {Lee},
  E.~{M{\"o}bius}, \& T.~H. {Zurbuchen}, 147--152

\bibitem[{{Van Hollebeke} {et~al.}(1978){Van Hollebeke}, {McDonald}, {Trainor},
  \& {von Rosenvinge}}]{vanh1978}
{Van Hollebeke}, M.~A.~I., {McDonald}, F.~B., {Trainor}, J.~H., \& {von
  Rosenvinge}, T.~T. 1978, \jgr, 83, 4723

\bibitem[{{Wijsen} {et~al.}(2023){Wijsen}, {Li}, {Ding}, {Lario}, {Poedts},
  {Filwett}, {Allen}, \& {Dayeh}}]{wij2023}
{Wijsen}, N., {Li}, G., {Ding}, Z., {et~al.} 2023, Journal of Geophysical
  Research (Space Physics), 128, e2022JA031203

\bibitem[{{Wijsen} {et~al.}(2021){Wijsen}, {Samara}, {Aran}, {Lario},
  {Pomoell}, \& {Poedts}}]{wij2021}
{Wijsen}, N., {Samara}, E., {Aran}, {\`A}., {et~al.} 2021, \apjl, 908, L26

\bibitem[{{Wimmer-Schweingruber} {et~al.}(2021){Wimmer-Schweingruber},
  {Janitzek}, {Pacheco}, {Cernuda}, {Espinosa Lara}, {G{\'o}mez-Herrero},
  {Mason}, {Allen}, {Xu}, {Carcaboso}, {Kollhoff}, {K{\"u}hl}, {Freiherr von
  Forstner}, {Berger}, {Rodriguez-Pacheco}, {Ho}, {Andrews}, {Angelini},
  {Aran}, {Boden}, {B{\"o}ttcher}, {Carrasco}, {Dresing}, {Eldrum}, {Elftmann},
  {Evans}, {Gevin}, {Hayes}, {Heber}, {Horbury}, {Kulkarni}, {Lario}, {Lees},
  {Limousin}, {Malandraki}, {Mart{\'\i}n}, {O'Brien}, {Prieto Mateo},
  {Ravanbakhsh}, {Rodriguez-Polo}, {S{\'a}nchez Prieto}, {Schlemm}, {Seifert},
  {Terasa}, {Tyagi}, {Vainio}, {Walsh}, \& {Yedla}}]{wimmer2021}
{Wimmer-Schweingruber}, R.~F., {Janitzek}, N.~P., {Pacheco}, D., {et~al.} 2021,
  \aap, 656, A22

\bibitem[{{Zank}(1999)}]{zank1999}
{Zank}, G.~P. 1999, \ssr, 89, 413

\bibitem[{{Zank} {et~al.}(2018){Zank}, {Adhikari}, {Zhao}, {Mostafavi},
  {Zirnstein}, \& {McComas}}]{zank2018}
{Zank}, G.~P., {Adhikari}, L., {Zhao}, L.~L., {et~al.} 2018, \apj, 869, 23

\bibitem[{{Zank} {et~al.}(2015){Zank}, {Hunana}, {Mostafavi}, {Le Roux}, {Li},
  {Webb}, {Khabarova}, {Cummings}, {Stone}, \& {Decker}}]{zank2015}
{Zank}, G.~P., {Hunana}, P., {Mostafavi}, P., {et~al.} 2015, \apj, 814, 137

\bibitem[{{Zhao} {et~al.}(2016){Zhao}, {Li}, {Mason}, {Cohen}, {Mewaldt},
  {Desai}, {Ebert}, \& {Al-Dayeh}}]{zhao2016}
{Zhao}, L., {Li}, G., {Mason}, G.~M., {et~al.} 2016, Research in Astronomy and
  Astrophysics, 16, 190

\bibitem[{{Zhuang} {et~al.}(2024){Zhuang}, {Lugaz}, {Al-Haddad}, {Scolini},
  {Farrugia}, {Regnault}, {Davies}, {Yu}, {Winslow}, \& {Galvin}}]{zhuang2024}
{Zhuang}, B., {Lugaz}, N., {Al-Haddad}, N., {et~al.} 2024, \aap, 682, A107

\end{thebibliography}

\appendix 

\section{Calculation of Interplanetary Shock Parameters}\label{app}

The calculation of the plasma parameters reported in Table \ref{tabPlasma} was done using the methods employed by the Heliospheric Shock Database, generated and maintained at the University of Helsinki. Many of the equations used can be found in the accompanying paper \citet{kil2015}, as well as the documentation provided on their website http://www.ipshocks.fi/documentation.

For the following calculations, fixed time intervals for the upstream and downstream regions of the shock were used for each spacecraft, and are denoted via the “u” and “d” subscripts, respectively. For the upstream region, the interval begins roughly 9 minutes before shock front measurement and ends 1 minute before the shock. For the downstream region, the interval begins 2 minutes after shock front measurement and extends to 10 minutes after the shock. These intervals were chosen to ensure that the shock ramp was not included in the parameter calculations, as well as to allow for an adequate amount of data points in each region of the shock, though the exact number of these points can vary based on the resolution of the spacecraft. As a result, the minimum number of data points allowed within an analysis interval was set at three, and the interval was allowed to extend if this minimum was not met. This was the case with the parameters calculated at ACE, whose intervals were extended to be 5-15 minutes before the shock for the upstream region and 5-15 minutes after the shock for the downstream region. Quantities that were averaged over these intervals are denoted by the “u” and “d” subscripts, respectively, within the equations presented. Additionally, the upstream values (parameters with the “up” superscript) are also averaged over the upstream time intervals.

The upstream sound speed is defined as
\begin{equation}
    C_s^{u p}=\left\langle c_s\right\rangle_{u p}=\left\langle\sqrt{\gamma k_B \frac{T_p+T_e}{m_p}}\right\rangle_{u p}
\end{equation}
where $\gamma$ is the polytropic index (5/3 following an assumption of an isotropic monatomic ideal gas) , $k_{B}$ is the Boltzmann Constant and $T_{P}$ is the proton temperature. The electron temperature, $T_{e}$, is defined as a function of the radial distance of each spacecraft from the Sun
\begin{equation}
T_e=T_e\left(R_{\mathrm{AU}}\right)=146277 \cdot R_{\mathrm{AU}}^{-0.664} \mathrm{~K}
\end{equation}

The upstream Alfven speed is 
\begin{equation}
    V_A^{u p}=\left\langle v_A\right\rangle_{u p}=\left\langle\frac{B}{\sqrt{\mu_0 N_p m_p}}\right\rangle_{u p}
\end{equation}
where B is the magnetic field magnitude, $\mu_{0}$ is the vacuum permeability, $N_{p}$ is the proton density, and $m_p$ is the proton mass.

The upstream magnetosonic speed is a combination of the Alfven and sound speed
\begin{equation}
    V_{m s}^{u p}=\left\langle v_{m s}\right\rangle_{u p}=\left\langle\sqrt{v_A^2+c_s^2}\right\rangle_{u p}
\end{equation}
and the upstream plasma beta, or the ratio between the plasma and magnetic pressure, is defined as
\begin{equation}
    \beta^{u p}=\langle\beta\rangle_{u p}=\left\langle\frac{2 \mu_0 k_B N_p\left(T_p+T_e\right)}{B^2}\right\rangle_{u p}
\end{equation}

The shock normal is computed using both magnetic field and velocity measurements from either spacecraft according to the mixed mode method (MD3, from \citet{abyun1976}). The use of this method does include some caveats, namely in the intervals chosen for the up and downstream plasma. These intervals have to be entirely separate from the shock layer, correspond to the actual upstream and downstream data points and leave out disturbances unrelated to the shock, as well as be long enough to remove the influence of turbulence and wave activity. 
\begin{equation}
    \hat{\mathbf{n}}= \pm \frac{\left(\mathbf{B}_{\text {d }}-\mathbf{B}_{\text{u}}\right) \times\left(\left(\mathbf{B}_{\text {d }}-\mathbf{B}_{\text {u }}\right) \times\left(\mathbf{V}_{\text {d }}-\mathbf{V}_{\text {u }}\right)\right)}{\left|\left(\mathbf{B}_{\text {d }}-\mathbf{B}_{\text {u }}\right) \times\left(\left(\mathbf{B}_{\text {d }}-\mathbf{B}_{\text {u }}\right) \times\left(\mathbf{V}_{\text {d }}-\mathbf{V}_{\text {u }}\right)\right)\right|}
\end{equation}
In the event that the velocity data is unavailable, as is the case at ACE, we instead  use the magnetic field coplanarity method  to calculate the shock normal\citep{colson1966}:
\begin{equation}
    \hat{\mathbf{n}}_{\mathbf{MC}}= \pm \frac{\left(\mathbf{B}_{\text {d }}-\mathbf{B}_{\text{u}}\right) \times\left(\mathbf{B}_{\text {d }} \times\mathbf{B}_{\text {u }}\right)}{\left|\left(\mathbf{B}_{\text {d }}-\mathbf{B}_{\text{u}}\right) \times\left(\mathbf{B}_{\text {d }} \times\mathbf{B}_{\text {u }}\right)\right|}
\end{equation}
For both methods, the sign of the shock normal is determined by the type of shock: positive for fast forward shocks and negative for fast reverse shocks. It is important to note that these two methods might result in varying signs for the radial and tangential component of the shock normal. However, this does not seem to significantly effect the parameters that depend on this value.

The shock speed is calculated in the reference frame of the spacecraft using the mass flux conservation over the shock
\begin{equation}
    V_{s h}=\left|\frac{\left[\rho_m \mathbf{V}\right]}{\left[\rho_m\right]} \cdot \hat{\mathbf{n}}\right|=\left|\frac{N_p^{\text {d }} \mathbf{V}_{\text {d }}-N_p^{\text {u}} \mathbf{V}_{\text {u}}}{N_p^{\text {d }}-N_p^{\text {u}}} \cdot \hat{\mathbf{n}}\right|
\end{equation}

The shock theta, the angle between the normal vector and the upstream magnetic field lines, is defined as
\begin{equation}
    \theta_{B n}=\frac{180^{\circ}}{\pi} \arccos \left(\frac{\left|\mathbf{B}_{\text {u}} \cdot \hat{\mathbf{n}}\right|}{\left\|\mathbf{B}_{\text {u}}\right\|\|\hat{\mathbf{n}}\|}\right)
\end{equation}

Finally, both the Alfven and Magnetosonic Mach numbers were calculated using a Galilean coordinate transformation to the rest frame of the shock, in which the solar wind velocity was transformed from $\mathbf{V}_{\text {u}}$ to $\mathbf{V}_{\text {u}}^{\prime}$. The Alfven mach number is defined as 
\begin{equation}
    M_A=\frac{\left|\mathbf{V}_{\text {u}}^{\prime} \cdot \hat{\mathbf{n}}\right|}{V_A^{\text {u}}}=\frac{\left|\mathbf{V}_{\text {u}} \cdot \hat{\mathbf{n}} \pm V_{s h}\right|}{V_A^{\text {u}}}
\end{equation}
and the magnetosonic mach number is defined as
\begin{equation}
    M_{m s}=\frac{\left|\mathbf{V}_{\text {u}}^{\prime} \cdot \hat{\mathbf{n}}\right|}{V_{m s}^{\text {u}}}=\frac{\left|\mathbf{V}_{\text {u}} \cdot \hat{\mathbf{n}} \pm V_{s h}\right|}{V_{m s}^{\text {u}}}
\end{equation}
where the sign of the shock velocity is also determined by the type of shock: negative for fast forward shocks and positive for fast reverse shocks. Both mach numbers are dependent upon the shock speed, shock normal, and the upstream Alfven and magnetosonic speeds, respectively. 

\end{document}